\documentclass[a4paper,fleqn,usenatbib]{mnras}
\pdfoutput=1

\usepackage{newtxtext,newtxmath}

\usepackage[T1]{fontenc}
\usepackage{ae,aecompl}

\usepackage{graphicx}	
\usepackage{amsmath}	
\usepackage{amssymb}	

\graphicspath{{Figures/}}

\title[An irregular discrete time series model]{An irregular  discrete time series model to identify residuals with autocorrelation in astronomical light curves}

\author[Eyheramendy et al.]{
Susana Eyheramendy,$^{1,2,3}$\thanks{E-mail: susana@mat.puc.cl}
Felipe Elorrieta,$^{1,2}$
Wilfredo Palma$^{1,2}$
\\
$^{1}$Department of Statistics, Faculty of
  Mathematics, Pontificia Universidad
  Cat\'olica de Chile, Av.\ Vicu\~na Mackenna
    4860\\, 7820436 Macul, Santiago, Chile\\
$^{2}$Millennium Institute of Astrophysics, Santiago, Chile\\
$^{3}$Max-Planck-Institut f\"ur Astronomie, Heidelberg, Germany
}

\date{Accepted XXX. Received YYY; in original form ZZZ}

\pubyear{2018}

\begin{document}
\label{firstpage}
\pagerange{\pageref{firstpage}--\pageref{lastpage}}
\maketitle

\begin{abstract}
Time series observations are ubiquitous in astronomy, and are generated  to distinguish between different types of supernovae, to detect and characterize extrasolar planets and to classify variable stars. 
These time series are usually modeled using a parametric and/or physical model that assumes independent and homoscedastic errors, but in many cases these assumptions are not accurate and there remains a temporal dependency structure on the errors. This can occur, for example,  when the proposed model cannot explain all the variability of the data or when the parameters of the model are not properly estimated.  In this work we define an autoregressive model for irregular discrete-time series, based on the discrete time representation of the continuous autoregressive model of order 1. We show that the model is ergodic and stationary. We further propose a maximum likelihood estimation procedure and assess the finite sample performance by Monte Carlo simulations.  We implement the model on real and simulated data from  Gaussian as well as other distributions, showing that the model can flexibly adapt to different data distributions.
We apply the irregular autoregressive model to the residuals of a transit of an extrasolar planet to illustrate errors that remain with temporal structure. We also apply this model to residuals of an harmonic fit of light-curves from variable stars to illustrate how the model can be used to detect incorrect parameter estimation.

\end{abstract}




\section{Introduction}
\label{sec:intro}

An irregular time series is a sequence of observational times and values $(t_n,y_n)$ such that the series $t_1,\ldots,t_N$ is strictly increasing, the distance between consecutive times, $t_j-t_{j-1}$ in general differs, and $y_1,\ldots,y_N$ is a sequence of real numbers.  Irregular time series are commonly observed in many  disciplines. For example, natural disasters, such as earthquakes, floods or volcano eruptions, occur with different time gaps. In the health science, patients can be  observed irregularly in time,   and  in astronomy observations are usually obtained at irregular time gaps due to, for example, its dependency on clear skies to be able to get observational data from optical telescopes. 

The analysis and modeling of time series are common and there exists a vast amount of theory and methods, most of which assume equally spaced measurements \citep[e.g.][]{Brockwell_1991,Brockwell_2016,Box_2015}. In practice, often  the analysis of irregularly spaced data is performed by ignoring the irregularity of the times and assuming regular spaced data. This practice can introduce bias in the parameter estimation  leading to inaccurate predictions. Another common practice is to transform the irregular time series into a regular time series by performing interpolation, usually linear, and then apply methodology for equally spaced data (see for example \citet{Adorf_1995} for a survey of such methods in the context of astronomical data). This again can introduce significant bias in the parameter estimation, specially when the time gap differences vary a lot (see for example \citet{Eckner_2014} for more details).

Some efforts have been made in trying to develop models for irregular time series. For example, \cite{Erdogan_2005}, introduces two models, one that assumes stationarity which can be seen as an extension of the classical autoregressive model of order one (AR(1)), while the second model does not assume stationarity, allowing some flexibility. \cite{Eckner_2014} attempts to develop a general framework for modeling irregular time series consistent with existing methods on equally spaced time series, but does not consider model specification and estimation. Other authors have suggested to embed irregular time series into continuous diffusion processes (e.g. \cite{Jones_1985}) and use the Kalman filter to estimate  the parameters and to carry out predictions (e.g \cite{Belcher_1994,Parzen_1984}).
  
  In astronomy considerable effort has been put in the estimation of spectrum of an irregular time series (e.g. \cite{Lomb_1975, Scargle_1982, Thiebaut_2005}), and some effort in the modeling. For example, \cite{Tuomi_etal2013} developed a first order autoregressive model, a first order moving average model and a general ARMA model, but these models do not have desirable statistical properties, as they are neither stationary\footnote{ A stationary process is a stochastic process whose unconditional joint probability distribution does not change when shifted in time.} nor ergodic\footnote{A stochastic process is said to be ergodic if its statistical properties can be deduced from a single, sufficiently long, random sample of the process.}. \cite{BailerJones_2011} developed Bayesian models for terrestrial impact cratering to assess periodicity on the impact ages, and generalize the models for different types of data in \cite{BailerJones_2012}. \cite{Kelly_2014} follows a different approach by proposing to use continuous-time autoregressive moving average (CARMA) models to fit irregular time series. 
  
  Other methods have been developed that attempt to estimate the autocorrelation of a time series, which in general do not depend on a model but estimate the autocorrelation directly from the data (for a review of such methods see e.g. \cite{Rehfeld_2011}). But in general, for fitting light curves for example, there are two main approaches followed by astronomers that account for irregular spaced time series. One is to use Monte Carlo simulations to forward model the periodogram as a function of a model power spectrum, and the other approach is to fit the light-curves in the time domain fitting usually Gaussian processes (e.g. \cite{Rasmussen06gaussianprocesses,Foreman_2017}). Both general methodologies can be computationally very expensive (e.g. \cite{ Kelly_2014,Kelly_2009,Brewer_2011,Done_1992,Emmanoulopoulos_2013,Uttley_2002}).

Exceptions can be found on models that can be represented as state space models, such as the CARMA($p,q$) models. These models overcome the computational burden by using Kalman filter to estimate the likelihood function.

 In this study we consider the continuous autoregressive model of order $1$, the so-called CAR(1) model or CARMA(1,0). Based on the discrete-time representation of this model, we define the {\it irregular autoregressive model} (IAR), derive its statistical properties  and develop statistical tests to assess the significance of the parameter of the model.  We further show that this discrete representation of the autoregressive model allows for Gaussian and non-Gaussian distributed data, leading to increase flexibility.
 
 
    We focus on applications of the IAR model in astronomy, but the model could be applied to any other field as well. Models for irregular time series are particularly relevant in astronomy as current and future  time domain optical surveys, such as SDSS Stripe 82 Supernova Survey (\cite{Frieman_2008}), Palomar Transient Factory (PTF, \cite{Law_2009}), the Catalina Real-Time Transient Survey (CRTS, \cite{Drake_2009}), Pan-STARRS (\cite{Kaiser_2002}), and the Large Synoptic Survey Telescope (LSST, \cite{Ivezic_2008}), will provide a huge amount of data in the form of irregular time series. 
    
    In this article, the models and its properties are shown in \S~\ref{sec:methods}. In \S~\ref{sec:car} the CAR(1) model is described, while in \S~\ref{sec:iar} the IAR model is defined and its statistical properties derived. We assess the finite sample performance of the maximum likelihood estimator of the parameter of the IAR model via Monte Carlo simulations and show the results in \S~\ref{sec:sim}. We compare the performance of the IAR model with the regular autoregressive model of order one and the ARFIMA models, and show the results in \S~\ref{sec:simRegular}. In \S~\ref{sec:gamma} we illustrate how the IAR model can fit a Gamma and a Student-t distributed sequence. Further, in order to illustrate some possible uses of this model in astronomy, we implement the IAR model in the context of two astronomical dataset (\S~\ref{sec:astro}). We implement the model on light-curves of variable stars obtained from the OGLE and Hipparcos surveys (\S~\ref{sec:variableStars}) and on a light-curve from a star with a transiting exoplanet (\S~\ref{sec:planet}). We develop statistical tests to assess the significance of the single parameter of the model, which allows to check whether there remain significant autocorrelation on the time series. We develop an algorithm for maximum likelihood estimation. We implement code in the R statistical software and Python to estimate the model and to perform the statistical test that assess significance. We end this paper with a discussion in \S~\ref{sec:discussion}. 

\section{Time series models and their properties}
\label{sec:methods}

We consider astronomical time series that can be fit using a parametric model that is represented  as  

\begin{equation}
z_t=g(t,\theta)+\delta_t,
\label{eq:general}
\end{equation}

\noindent
where $z_t$ is the astronomical observation at time $t$, $g(t,\theta)$ is the mean of the model at time $t$, that depends on the vector of parameters $\theta$, and $\delta_t$ is the error of the model at time $t$. 

For example, in fitting light-curves of periodic variable stars, the usual approach is to use an harmonic model where 
\begin{equation}
\label{eq:harmonic}
 g(t,\theta)=\alpha+\beta t+\mathop{\sum}\limits_{j=1}^4 (a_{j} \mbox{sin}(2\pi f_1 jt) +  b_{j} \mbox{cos}(2\pi f_1 jt))
  \end{equation}
  
\noindent and $\theta=(\alpha,\beta,f_1,a_1,\ldots,a_4,b_1,\ldots,b_4)$.  In this case, $z_t$ represents the flux measurement of the variable star at time $t$, $\alpha$ and $\beta$ are the parameters of a linear trend, $1/f_1$ is the period  of the star, $\{a_j\}$ and $\{b_j\}$ are the parameters of the harmonic model. For transient or variable phenomena, such as supernovae or planets, $g(t,\theta)$ is fit using a deterministic statistical or astrophysical model.

These errors (i.e. \{$\delta_t$\}) are usually assumed independent with a Gaussian distribution with mean zero and variance $\sigma^2$. In many cases neither the independence of the errors nor the homoscedasticity (or equal variance) of the errors is achieved.  To identify and overcome these problems, the continuous autoregressive model (CAR), for example,  can be implemented on  $z_t-g(t,\hat{\theta})$, i.e. the residuals of the model in equation (\ref{eq:general}), in order to assess whether a correlation structure remains after fitting such model.

 In  the following two sections we describe the CAR(1) model and define the irregular autoregressive (IAR) model. These models would typically be used  to identify autocorrelation in the residuals.

\subsection{Continuous autoregressive model of order 1}
\label{sec:car}
The continuous autoregressive model  of order $1$ (CAR(1)) attempts to solve a stochastic differential equation of order one, driven by white noise. White noise is the name used in time series analysis for an independent series of random variables (when the data is assumed to be Gaussian). The problem is that continuous time white noise exists  only in the sense that its integral is a continuous time random walk, commonly referred as Brownian motion or Wiener process. A continuous time random walk is the limit of a discrete time random walk as the time interval gets small. The path function of a Wiener process can be simulated, and will be continuous  with a very wiggly appearance and its derivative does not exist. Moreover, a finite segment of this curve has infinite path length. Despite all these undesirable properties the Wiener process is still the key to get random input into a continuous time process (\cite{Jones_1993}). 

The mathematical formulation of the process $\epsilon (t)$ corresponding to a CAR(1) model is 

\begin{equation}
\label{eq:car1}
\frac{d}{dt}\epsilon (t)+\alpha_0 \epsilon (t)=\sigma_0 \nu (t) +\beta,
\end{equation}

\noindent
where $\nu (t)$ is the continuous time white noise, and $\alpha_0$ and $\beta$ are unknown parameters of the model.
It can be shown that the process $\epsilon(t_k)$ that is a solution of  (\ref{eq:car1}), is  also a solution of the difference equation given by

\begin{equation}
\label{eq:car1iar}
\epsilon (t)-\frac{\beta}{\alpha_0}=e^{-\alpha_0(t-s)}(\epsilon(s)-\frac{\beta}{\alpha_0})+e^{-\alpha_0 t}(I(t)-I(s))
\end{equation}

\noindent
where $I(t)=\sigma_0\int_0^te^{\alpha_0u}dw(u)$ is an It$\hat{\mbox{o}}$ integral\footnote{The integral is an extension of the Riemann-Stieltjes integral, where the integrands and the integrators are now stochastic processes.} (\cite{Brockwell_2016}). See Appendix~\ref{sec:car1} for a full derivation of this result. Based on this last equation, we define the discrete time series model for irregularly sampled observations and derive its statistical properties (shown on the following sections).

\subsection{Irregular Autoregressive (IAR) model}
\label{sec:iar}
Denote $y_{t_j}$ an observation measured at time $t_j$, and consider an increasing sequence of observational times $\{t_j\}$ for $j=1,\dots,n$. We define the irregular autoregressive (IAR) process by
\begin{equation}  \label{Model} 
 y_{t_j}=\phi^{t_j-t_{j-1}} \, y_{t_{j-1}} + \sigma \, \sqrt{1-\phi^{2(t_j-t_{j-1})}}  \, \varepsilon_{t_j} 
\end{equation}

\noindent
where $\varepsilon_{t_j}$ are independent random variables with zero mean and unit variance. Note that by replacing $ e^{-\alpha_0}$ with $\phi$ (and setting $\beta=0$) in (\ref{eq:car1iar}) we get to the Gaussian IAR model, because the CAR(1) model assumes Gaussian data. The connection between equations \eqref{eq:car1iar} and \eqref{Model} are completed by defining $\sigma^2 = \frac{\sigma_0^2}{2 \alpha_0}$. 

Importantly, the model described by Equation \eqref{Model} can also be established without assuming Gaussian errors. From now on, we do not assume Gaussian data  to derive the statistical properties of the model unless we explicitly mention a distribution assumption.
  
Observe that
\begin{equation}
E(y_{t_j})=0 \mbox{ and }  Var(y_{t_j})=\sigma^2 \mbox{ for all } y_{t_j}, 
\end{equation}

\noindent
and the covariance between $y_{t_k}$ and $y_{t_j}$ is  $E(y_{t_k} \, y_{t_j})= \sigma^2 \,  \phi^{t_k-t_j}$, for $k\geq j.$

\noindent
 Thus, for any two observational times $s<t$ we can define the autocovariance function as
\begin{equation}
\gamma(t-s)=E(y_t \, y_s)= \sigma^2 \,  \phi^{t-s},
\end{equation}
as well as the autocorrelation function (ACF), $\rho(t-s)=\frac{\gamma(t-s)}{\gamma(0)}= \phi^{t-s}$.

Given the results above, the sequence $\{y_{t_j}\}$ corresponds to a second-order or weakly stationary\footnote{A weakly stationary process is a random sequence of random variables that requires that the first moment (i.e. the mean) and the autocovariance do not vary with respect to time.} process. We show in the next theorem that, in addition, under some conditions the process is stationary and ergodic.

\vspace{0.1in}

{\bf Theorem 1: }Consider the process defined by (\ref{Model}) and assume that the input noise is an i.i.d. sequence of random variables with zero mean and unit variance. Furthermore, suppose that  $t_j-t_{j-n} \geq C\log n$ as $n\to \infty$, $0< \phi<1$ where $C$ is a positive constant  that satisfies $C \log \phi^2<-1$. Then, there exists a solution to the process defined by  (\ref{Model}), and the sequence $\{y_{t_j}\}$ is stationary and ergodic. See Appendix~\ref{sec:teo1} for a proof of this theorem.\\

Note that, if $t_j-t_{j-1}=1$ for all $j$, then equation (\ref{Model}) becomes 

\begin{equation} 
y_{t_j}=\phi \, y_{t_{j-1}} + \sigma \, \sqrt{1-\phi^2}  \, \varepsilon_{t_j} \hspace{0.1cm}\mbox{ for } j=2,\ldots,n,
\end{equation}

\noindent
which corresponds to the autoregressive model of order $1$ (AR(1)) for regularly space data. Therefore the IAR model is an extension of the regular autoregressive model. As mentioned previously, it is also an extension of the continuous autoregressive model of orden $1$.

Note also that for the regular AR(1) model the two assumptions on the theorem are satisfied: $t_j-t_{j-n}=n$, $n>\mbox{log} (n)$ is achieved, and $\phi^2<1$ is part of the assumptions of the regular autoregressive model. Therefore the AR(1) is ergodic and stationary. 
\vspace{0.2in}

{\bf Corollary:} Let $\bar{y}_n=\frac{1}{n}\sum _{j=1}^ny_{t_j}$ and $\tilde{\sigma}_n^2  =\frac{1}{n}\sum_{j=1}^n    (y_{t_j} -\bar{y}_n     )^2$ be the sample mean and the sample variance of the IAR process, respectively. Then, we have that
$\bar{y}_n \to E(y_{t_j}) $ and $\tilde{\sigma}_n^2  \to \sigma^2$,
in probability, as $n\to\infty$.

\subsection{Estimation}
The likelihood of the data $\{y_{t_1},\ldots,y_{t_n}\}$ can be expressed as

\begin{equation}\label{eq:lik}
f(y_{t_1},\ldots,y_{t_n};\theta)=f(y_{t_1};\theta)f(y_{t_2}|y_{t_1};\theta)\times \ldots \times f(y_{t_n}|y_{t_{n-1}};\theta),
\end{equation}

\noindent
where  $\theta=(\sigma^2,\phi)$ is the parameter vector of the model. To describe clearly the estimation process, we assume here that the marginal and conditional distributions of the time series are Gaussian. Note that this assumption is not necessary to obtain the statistical properties stated in Theorem~$1$. In Section~\ref{sec:gamma} we show an example where the conditional distribution is assumed to be Gamma, and in Section~\ref{sec:std} we show an example where the conditional distribution is assumed to be a Student-t.

Assume that, 
\begin{equation}\label{eq:lik1}
f(y_{t_1};\sigma^2,\phi)\sim N(0,\sigma^2) \mbox{ and }
\end{equation}

\begin{equation} \label{eq:lik2}
f(y_{t_j}|y_{t_{j-1}};\sigma^2,\phi)\sim N(\phi^{t_j-t_{j-1}} \, y_{t_{j-1}} , \sigma^2 \, (1-\phi^{2(t_j-t_{j-1})})
\end{equation}
for $j=2,\dots ,n$. Based on equation~(\ref{Model}), minus the log-likelihood of this process can be written as
\begin{equation}\label{eq:loglik}
\ell(\theta)=\frac{n}{2}\log (2\pi)+\frac{1}{2}\sum_{j=1}^n \log \nu_{t_j} + \frac{1}{2}\sum_{j=1}^n \frac{e_{t_j}^2}{\nu_{t_j}}, 
\end{equation}

\noindent
where we define   $e_{t_1}=y_{t_1}$,  $e_{t_j}=y_{t_j}-\phi^{t_j-t_{j-1}} \, y_{t_{j-1}}\mbox{ for }j>1$ and their variances as $\nu_{t_j}=Var(e_{t_j}).$

Observe that the finite past predictor of the process at time $t_j$ is given by
\begin{equation}
\widehat{y}_{t_1}=0, \mbox{ and }\widehat{y}_{t_j}=\phi^{t_j-t_{j-1}} \, y_{t_{j-1}}, \mbox{ for }j=2,\dots,n. 
\end{equation}

Therefore, $e_{t_j}=y_{t_j}-\widehat{y}_{t_j}$ is the  prediction error  with variance $\nu_{t_1}=Var(e_{t_1})=\sigma^2$,

\begin{equation}
\nu_{t_j}=Var(e_{t_j})=\sigma^2 [1-\phi^{2(t_j-t_{j-1})}], \mbox{ for }j=2,\dots,n. 
\label{eq:nu}
\end{equation}

By direct maximization of the log-likelihood  (\ref{eq:loglik}),  we can obtain the maximum likelihood estimator of $\sigma^2$,

\begin{equation}
\hat{\sigma}^2=\frac{1}{n}\sum_{j=1}^n\frac{(y_{t_j}-\widehat{y}_{t_j})^2}{\tau_{t_j}}, \mbox{ where }\tau_{t_j}=\nu_{t_j}/\sigma^2.
\end{equation}

But it is not possible to find $\widehat{\phi}$, the maximum likelihood estimator of $\phi$,  by direct maximization of the likelihood, but iterative methods can be used (for details see Chapter~5 of \cite{Palma_16}). We developed scripts in the statistical language/software $R$ , and also in Python,  to estimate $\phi$.

{\bf Lemma 1:} Consider the process defined by (\ref{Model}) and  suppose that  $t_j-t_{j-n} =  h \, n$, for a positive constant $h$, $0< \phi<1$. Let $\widehat{\phi}_n$ be the maximum likelihood estimator of $\phi$. Then, the MLE satisfies the following asymptotic normal distribution:
\begin{equation}
\sqrt{n} \, (\widehat{\phi}_n-\phi) \to \rm{N}(0,\sigma_{\phi}^2),
\end{equation}
as $n \to \infty$, where 
\begin{equation}
\sigma_{\phi}^2 = \frac{1-\phi^{2h}}{h^2 \phi^{2h-2}}.
\end{equation}
 See Appendix~\ref{sec:lem1} for a proof of this lemma.\\

Similar to the continuous time autoregressive models, the IAR can be represented using state-space models from which the Kalman filter (\cite{kalman_1960}) can be implemented allowing fast and scalable estimation of parameters. 

\section{Simulation study to assess the maximum likelihood estimators of the IAR model}
\label{sec:sim}

This section shows the results of Monte Carlo experiments assessing the finite sample performance of the proposed maximum likelihood estimator. 

The simulated processes correspond to the model (\ref{Model}) where the observational times follow a mixture of two exponential distributions with means $1/\lambda_1$ and $1/\lambda_2$ respectively, and random weights $w_1$ and $w_2$, respectively. We find that this choice for the observational times corresponds to a reasonable representation for the observational times of a multi-year large time series survey such as the Vista Variable of the Via Lactea (\cite{VVV}). Table~\ref{tab:sim1} shows a summary of the simulations based on 1000 repetitions with  $\lambda_1=130$, $\lambda_2=6.5$, $w_1=0.15$ and $w_2=0.85$.  Table \ref{tab:sim2} shows a summary of the simulations based on 1000 repetitions with  $\lambda_1=300$, $\lambda_2=10$, $w_1=0.15$ and $w_2=0.85$.

\begin{table}
\centering
\caption{\em Maximum likelihood estimation of simulated IAR series with mixture of Exponential distribution for the  observational times, with $\lambda_1=130$ and $\lambda_2=6.5$, $w_1=0.15$ and $w_2=0.85$. \label{tab:sim1}}
\begin{tabular}{rrrrrrr}
  \hline
 Case & n & $\phi$ &  $\widehat{\phi}$ & SD($\widehat{\phi})$ & $\sigma(\widehat{\phi})$ &  $\widehat{\sigma}$\\ 
  \hline
1 & 50 & 0.900 & 0.887 & 0.044 & 0.034 & 1.013 \\
  2 & 50 & 0.990 & 0.985 & 0.008 & 0.008 & 1.039 \\
  3 & 50 & 0.999 & 0.996 & 0.004 & 0.003 & 1.155 \\
  4 & 100 & 0.900 & 0.894 & 0.029 & 0.024 & 1.005 \\
  5 & 100 & 0.990 & 0.988 & 0.005 & 0.006 & 1.015 \\
  6 & 100 & 0.999 & 0.998 & 0.002 & 0.002 & 1.049 \\
    \hline
\end{tabular}
\end{table}

\begin{table}
\centering
\caption{\em Maximum likelihood estimation of simulated IAR series of size $n$, with Exponential distribution mix observation times, $\lambda_1=300$ and $\lambda_2=10$. \label{tab:sim2}}
\begin{tabular}{rrrrrrr}
  \hline
 Case & n & $\phi$ &  $\widehat{\phi}$ & SD($\widehat{\phi})$ & $\sigma(\widehat{\phi})$ &   $\widehat{\sigma}$\\ 
  \hline
1 & 40& 0.900 & 0.8843 & 0.058 & 0.038 & 1.011 \\
  2 & 40 & 0.990 & 0.9854 & 0.009 & 0.007 & 1.037 \\
  3 & 40& 0.999 & 0.9969 & 0.003 & 0.002 & 1.120 \\
  4 & 80 & 0.900 & 0.8929 & 0.034 & 0.027 & 1.006 \\
  5 & 80 & 0.990 & 0.9876 & 0.005 & 0.005 & 1.018 \\
  6 & 80 & 0.999 & 0.9980 & 0.001 & 0.002 & 1.046 \\
   \hline
\end{tabular}
\end{table}

The Monte Carlo simulations suggest that the finite-sample performance of the proposed methodology is accurate. In particular, the estimation bias is small even for the smaller sample sizes used in Table~\ref{tab:sim1} and \ref{tab:sim2}. Note that we restrict to high values of the parameter $\phi$. The reason for this is the choice of the distribution of the observational time gaps, which tend to be large. Observe that an approximate asymptotic estimation of the standard deviation $\sigma(\widehat{\phi})$ obtained by an application of Lemma~$1$ is also provided in these tables. Notice that the approximation  seems to work well for larger sample sizes (e.g. 80 or 100)  and high values of $\phi$ (e.g. 0.9900 or 0.9990).\\

To assess whether the observational time distribution has any effect on the parameter estimation, we perform another Monte Carlo experiment using a quasi-periodic distribution.  To generate a sample of size $n$ of these times we use the following scheme. First, we assume a year of 365 days and then we randomly select $\frac{n}{10}$ observations from the uniform distribution $U(180,210)$ and another $\frac{n}{10}$ observations from the uniform distribution $U(240,270)$. This is repeated for five consecutive years. In this way, we obtain $n$ observational times at two fixed months a year (June and August), but on randomly picked days within the month. 

The finite sample performance is assessed by a simulation experiment based on $1000$ repetitions of the IAR process of sizes $n=60$ and $n=100$. The observational times are generated using the procedure mentioned above. Comparing the results in Tables \ref{tab:sim1} and \ref{tab:sim3}, we can conclude that the accuracy of the proposed estimation method is not affected by a quasi-periodic sample of the observational times.

\begin{table}
\centering
\caption{\em Maximum likelihood estimation of simulated IAR series with quasi periodic behavior in the observational times. \label{tab:sim3}}
\begin{tabular}{rrrrrr}
  \hline
Case & n & $\phi$ &  $\widehat{\phi}$ & SD($\widehat{\phi})$ &  $\widehat{\sigma}$\\ 
  \hline
1 & 60 & 0.900 & 0.887 & 0.039 & 1.011 \\
  2 & 60 & 0.990 & 0.985 & 0.008 & 1.022 \\
  3 & 60 & 0.999 & 0.996 & 0.003 & 1.108 \\
  4 & 100 & 0.900 & 0.890 & 0.032 & 1.009 \\
  5 & 100 & 0.990 & 0.986 & 0.008 & 1.013 \\
  6 & 100 & 0.999 & 0.996 & 0.003 & 1.076 \\
   \hline
\end{tabular}
\end{table}

\section{Simulation study to compare the IAR model with other models for regular time series}
\label{sec:simRegular}

\begin{figure*}
\centering
\includegraphics[width=\textwidth]{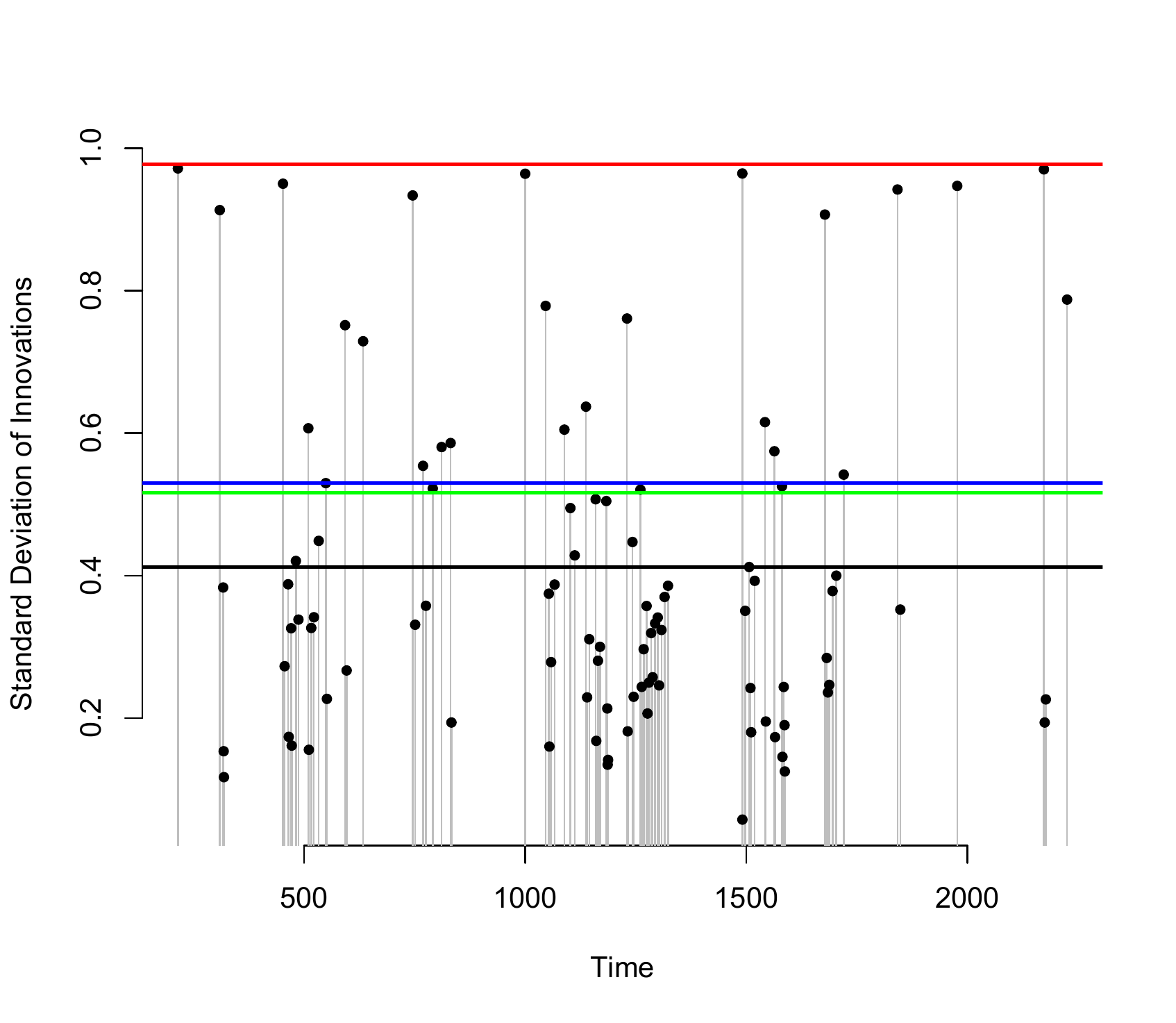}
\caption{Comparison of standard deviation at each time of a sequence simulated with the IAR model with parameter $\phi=0.99$ and length $100$. The red line corresponds to the standard deviation of the sequence, the blue and green lines correspond to the standard deviation estimated assuming an AR(1) and ARFIMA(1,d,0) model respectively. The black line corresponds to the average standard deviation of the IAR model, where the black dots are the individual standard deviations at each time.\label{fig:simIAR}}
\end{figure*}

We compare the IAR model with other standard models for regular time series.   Figure~\ref{fig:simIAR} shows the standard deviation of the prediction errors, i.e. the root of the series $\hat{\nu}_t$ in equation (\ref{eq:nu}). Note that because in the IAR model the prediction errors $e_{t_j}$ are unbiased, i.e. $E(e_{t_j})=0$,  the standard deviation of the prediction errors are equivalent to the root mean squared error (RMSE).

To estimate the prediction errors  we generate the sequence $\{y_1,\ldots,y_n\}$ using the IAR model with $\phi=0.99$ and $n=100$. The red line corresponds to the standard deviation of the sequence, the blue and green line correspond to the standard deviation of the regular autoregressive model of order one (AR(1)) and ARFIMA(1,d,0) respectively. These models assume regular spaced data.  The observational times are generated using the density $f(t|p,\lambda_1,\lambda_2)=p\mathcal{E}(t|\lambda_1)+(1-p)\mathcal{E}(t|\lambda_2)$ with $p=0.15, \lambda_1=130,\lambda_2=6.5$, where $\mathcal{E}(t|\lambda_1)$ denotes an Exponential distribution with parameter $\lambda_1$.

Observe that the only model that changes the standard deviation at each observational time is the IAR model, corresponding to the black dots in Figure~\ref{fig:simIAR}, where larger values close to one are observed after a larger observational time gap. The average standard deviation of the IAR model is shown as the black line, and it is smaller than the standard deviation of any of the other models.

In the next two sections we show  simulation studies to describe how the IAR model can be implemented to fit Gamma and Student-t distributed series and compare its performance with the continuous autoregressive model.


\section{Implementation of the IAR model on simulated Gamma distributed series}
\label{sec:gamma}

We implement the IAR model on simulated conditional Gamma distributions following the procedure described at \cite{Palma_11}. Specifically,  the conditional mean and variance of the IAR model are defined as

\begin{equation} 
\begin{split}
\mathbb{E}(y_{t_j}|y_{t_{j-1}}) &=& \mu + \phi^{t_j-t_{j-1}} \, y_{t_{j-1}} \\
\mathbb{V}(y_{t_j}|y_{t_{j-1}}) &=& \sigma^2 \, (1-\phi^{2(t_j-t_{j-1})}).
\end{split}
\label{eq:condIAR}
\end{equation}

These moments are equivalent to the ones for the Gaussian case stated in equation (\ref{eq:lik2}), the only difference is the positive parameter $\mu$ that corresponds to the expected value of $y_{t_{j-1}}$. If $y_{t_j}|y_{t_{j-1}}$ follows a Gamma distribution, a positive value of $\mu$ is required in order to ensure that the process is positive. However, the process may be shifted, so that $y_{t_j} - \mu$ have zero mean, like the Gaussian IAR. For simplicity, we set $\mu=1$.

In addition, note that under the assumption of stochastic times the marginal mean $\mathbb{E}(y_{t_j}) = \frac{\mu}{1-\mathbb{E}(\phi^{t_j-t_{j-1}})}$ and marginal variance $\mathbb{V}(y_{t_j}) = \sigma^2 + \frac{\mathbb{E}(y_{t_j})^2\mathbb{V}(\phi^{t_j-t_{j-1}})}{1-\mathbb{E}(\phi^{2(t_j-t_{j-1})})}$ are constants.
  
If $x_{t_j}\sim$~Gamma($\alpha_{t_j}$,$\beta_{t_j}$) follows a Gamma distribution with shape $\alpha_{t_j}$ and  scale $\beta_{t_j}$ parameters, it is well known that the expected value and the variance of $x_{t_j}$ are $\mathbb{E}(x_{t_j}) = \alpha_{t_j} \, \beta_{t_j}$  and $\mathbb{V}(x_{t_j}) = \mathbb{E}(x_{t_j}) \, \beta_{t_j}$ respectively. From the two equations,
\begin{eqnarray*} 
 \alpha_{t_j} \, \beta_{t_j}& = & \mu + \phi^{t_j-t_{j-1}} \, y_{t_{j-1}} \\
 \alpha_{t_j} \, \beta_{t_j}^2& = &\sigma^2 \, (1-\phi^{2(t_j-t_{j-1})}),
\end{eqnarray*} 
\noindent
we obtain, $\alpha_{t_j}$ and  $ \beta_{t_j}$ as functions of the parameters $\phi$ and $\sigma^2$: $\alpha_{t_j}= \alpha_{t_j}(\phi,\sigma^2) $ and  $ \beta_{t_j}= \beta_{t_j}(\phi,\sigma^2) $. Thus, the log-likelihood of the conditional distribution of $y_{t_j}|y_{t_{j-1}}$ can be written as,

\begin{eqnarray*} 
\ell_j &=& \log f_{\theta}\left(\phi,\sigma^2\right) \\
&=& - \left(\alpha_{t_j}\right) \log \beta_{t_j} - \log \Gamma \left(\alpha_{t_j}\right) - \frac{1}{\beta_{t_j}}y_{t_j} + \left(\alpha_{t_j}-1\right) \log y_{t_j} 
\end{eqnarray*} 

Here we omit the dependence of $ \alpha_{t_j}$ and $ \beta_{t_j}$ on $\phi$ and $\sigma^2$ to keep notation clear.
If $y_{t_1} \sim {\rm Gamma}(1,1)$, then the full log-likelihood is,

$$\ell(\theta) = \mathop{\sum}\limits_{j=2}^{N} \ell_j + \ell_1$$

\noindent
where $\ell_1 = - y_{t_1}$. The unknown parameters of the model are $\phi$ and $\sigma$ which can be estimated using iterative methods.\\

We perform Monte Carlo experiments, based in $1000$ repetitions, and we assess the accuracy in parameter estimation on simulated conditionally Gamma distributed time series. We implement the Gamma distributed IAR model as well as the Gaussian distributed IAR model in the statistical software package R and Python. The Gaussian distributed IAR model (i.e., samples from a CAR(1) model) is implemented using the R package \texttt{cts} and the Python script developed by \cite{Pichara_2012}.

In Table~\ref{tab:gammaR}, $\widehat{\phi}$ corresponds to the estimator using the correct Gamma distributed and $\widehat{\phi}^C$  is the estimator using the mismatched Gaussian distributed IAR model. The performance of the Gaussian distributed IAR model using Python and R does not vary significantly. In both cases, performance assuming the mismatched Gaussian distributed IAR model is substantially inferior to assuming the correct Gamma distributed IAR model.

\begin{table*}
\centering
\caption{\em Implementation of  the Gamma distributed IAR model and the CAR(1) model on simulated Gamma-IAR series in  \textbf{R} and Python. For the observational times we use a mixture of two Exponential distributions with parameters  $\lambda_1=130$ and $\lambda_2=6.5$, $w_1=0.15$ and $w_2=0.85$. \label{tab:gammaR}}
\begin{tabular}{lrrrrrrrrr}
  \hline
 & N & $\phi$ & $\sigma$ & $\widehat{\phi}$ & SD($\widehat{\phi})$ & $\widehat{\phi}^C$ & SD($\widehat{\phi}^C)$ & $\widehat{\sigma}$  & SD($\widehat{\sigma})$ \\
  \hline
R & 100 & 0.9 & 1 & 0.899 & 0.014 & 0.418 & 0.306 & 0.984 & 0.170 \\
R & 100 & 0.99 & 1 & 0.990 & 0.001 & 0.890 & 0.201 & 0.985 & 0.161 \\
R & 200 & 0.9 & 1 & 0.899 & 0.010 & 0.355 & 0.286 & 0.993 & 0.122 \\
R & 200 & 0.99 & 1 & 0.990 & 0.001 & 0.900 & 0.184 & 0.998 & 0.120 \\
    \hline
Python & 100  & 0.9 & 1 & 0.899 & 0.013 & 0.449 & 0.318 & 0.990 & 0.169 \\
Python & 100 & 0.99  & 1 & 0.990 & 0.001 & 0.919 & 0.169 & 0.981 & 0.200 \\
Python & 200 & 0.9 & 1 & 0.899 & 0.010 & 0.393 & 0.299 & 0.985 & 0.127 \\
Python & 200 & 0.99 & 1 & 0.990 & 0.001 & 0.927 & 0.163 & 0.996 & 0.332 \\
    \hline
\end{tabular}
\end{table*}

\section{Implementation of the IAR model on Student-t distributed  series}
\label{sec:std}

Another implementation of a non-Gaussian IAR process is on a heavy-tailed distribution such as  the Student-t distribution. This kind of distribution are useful to address the problem of possible outliers in a time series. Following the procedure mentioned in Section \ref{sec:gamma}, we implement an IAR model with a Student-t conditional distribution. If $x_{t_j}\sim t_{\nu}(\lambda_{t_j},\tau^2_{t_j})$ follows a non-standardized Student's t-distribution with mean $\lambda_{t_j}$, variance $\tau^2_{t_j}$ and $\nu$ degrees of freedom, the expected value of $x_{t_j}$ is $\mathbb{E}(x_{t_j}) = \lambda_{t_j}$ and the variance is $\mathbb{V}(x_{t_j})=\tau_{t_j}^2 \frac{\nu}{\nu-2}$.  From the conditional mean and variance of IAR model \eqref{eq:condIAR} we  obtain,
\begin{eqnarray*} 
 \lambda_{t_j} &= & \phi^{t_j-t_{j-1}} \, y_{t_{j-1}} \\
  \tau^2_{t_j} & = &\frac{\nu-2}{\nu} \,\sigma^2\,\left(1-\phi^{2(t_j-t_{j-1})} \right).
\end{eqnarray*} 

Thus the log-likelihood of the conditional distribution of $y_{t_j}|y_{t_{j-1}}$ can be written as,

\begin{eqnarray*} 
\ell_j &=& \log f_{\nu}\left(\lambda_{t_j},\tau^2_{t_j}\right) \\
&=& \log \left(\frac{\Gamma \left( \frac{\nu+1}{2} \right)}{\Gamma \left(\frac{\nu}{2} \right)\sqrt{\nu \pi}}\right) \\
& & \hspace{0.1in} - \frac{1}{2}  \log \tau_{t_j}^2 -\frac{\nu+1}{2}  \log    \left( 1 + \frac{1}{\nu} \left(\frac{y_{t_j}-\lambda_{t_j}}{\tau_{t_j}} \right)^2 \right)
\end{eqnarray*} 

Let $y_{t_1} \sim N(0,1)$, then the full log-likelihood is,

$$\ell(\theta) = \mathop{\sum}\limits_{j=2}^{N} \ell_j + \ell_1$$

\noindent
where $\ell_1 =  - \frac{1}{2} (\log (2 \pi) + y_{t_1}^2)$.\\

In order to assess the accuracy in the parameter estimation procedure we also perform Monte Carlo experiments with $1000$ repetitions. We use two different values for the degrees of freedom $\nu =3$ and $\nu= 5$. Table \ref{tab:tR} shows that the estimation of the parameters $\phi$ and $\sigma$ is precise in both cases. As expected, the estimation performance of the Gaussian IAR model is similar to the one obtained with the Student-t distribution model.

\begin{table*}
\centering
\caption{\em Implementation of the T distributed IAR model and the CAR(1) model on simulated Student-t IAR series. For the observational times we use a mixture of two Exponential distributions with parameters  $\lambda_1=130$ and $\lambda_2=6.5$, $w_1=0.15$ and $w_2=0.85$. \label{tab:tR}}
\begin{tabular}{rrrrrrrrr}
  \hline
N & $\phi$ & $\nu$ & $\widehat{\phi}$ & SD($\widehat{\phi})$ & $\widehat{\phi}^C$ & SD($\widehat{\phi}^C)$ & $\widehat{\sigma}$  & SD($\widehat{\sigma})$ \\
  \hline
100 & 0.9 & 3 & 0.895 & 0.025 & 0.884 & 0.068 & 1.010 & 0.231 \\
100 & 0.99 & 3 & 0.988 & 0.005 & 0.983 & 0.045 & 0.979 & 0.360 \\
200 & 0.9 & 3 & 0.898 & 0.016 & 0.889 & 0.054 & 1.003 & 0.163 \\
200 & 0.99 & 3 & 0.989 & 0.003 & 0.987 & 0.005 & 0.991 & 0.258 \\
    \hline
100 & 0.9 & 5 & 0.896 & 0.028 & 0.892 & 0.037 & 1.010 & 0.225 \\
100 & 0.99  & 5 & 0.989 & 0.005 & 0.986 & 0.005 & 1.017 & 0.395 \\
200 & 0.9 & 5 & 0.897 & 0.018 & 0.895 & 0.023 & 1.006 & 0.157 \\
200 & 0.99 & 5 & 0.989 & 0.003 & 0.988 & 0.003 & 1.007 & 0.274 \\
    \hline
\end{tabular}
\end{table*}

\section{Examples of the  IAR model in Astronomical time series}
\label{sec:astro}
In this section we illustrate two implementation of the IAR model in Astronomical time series. The first implementation is to detect model misspecification, i.e. a model with incorrectly estimated parameters or that is not sufficiently complex to describe the data at hand. The second implementation is to identify the presence of time-correlated structure in model residuals. For the model misspecification case we use variable star light-curves from the OGLE and Hipparcos survey, and for the time-correlation structure we use a light curve of an exoplanetary transit.

\subsection{Application to variable stars from the OGLE and Hipparcos surveys}
\label{sec:variableStars}

The harmonic model described in equation (\ref{eq:harmonic}) is used to model light-curves from variable stars. This model requires first to find the period of the variable star, which can be estimated, for example, using the Generalized Lomb-Scargle periodogram \cite{Zechmeister_etal09}. Then the remaining parameters are estimated using techniques for maximizing the likelihood. For more details on the procedure of the modeling of periodic light-curve, see for example \cite{Debosscher_etal07},  \cite{Richards_etal11} or \cite{Elorrieta_etal16}.

 Denote the residuals after subtracting a linear trend and an harmonic model with one frequency and four components as $y_t$, i.e 
 
 \[y_t=z_t-\hat{\alpha}-\hat{\beta} t-\mathop{\sum}\limits_{j=1}^4 (\hat{a}_{j} \mbox{sin}(2\pi f_1 jt) +  \hat{b}_{j} \mbox{cos}(2\pi f_1 jt)),\]
 
 \noindent
 where $\hat{a}$ represents a maximum likelihood estimator. We implement the IAR model on these residuals.

\begin{center}
\begin{figure*}
\begin{minipage}{0.32\linewidth}
\includegraphics[width=\textwidth]{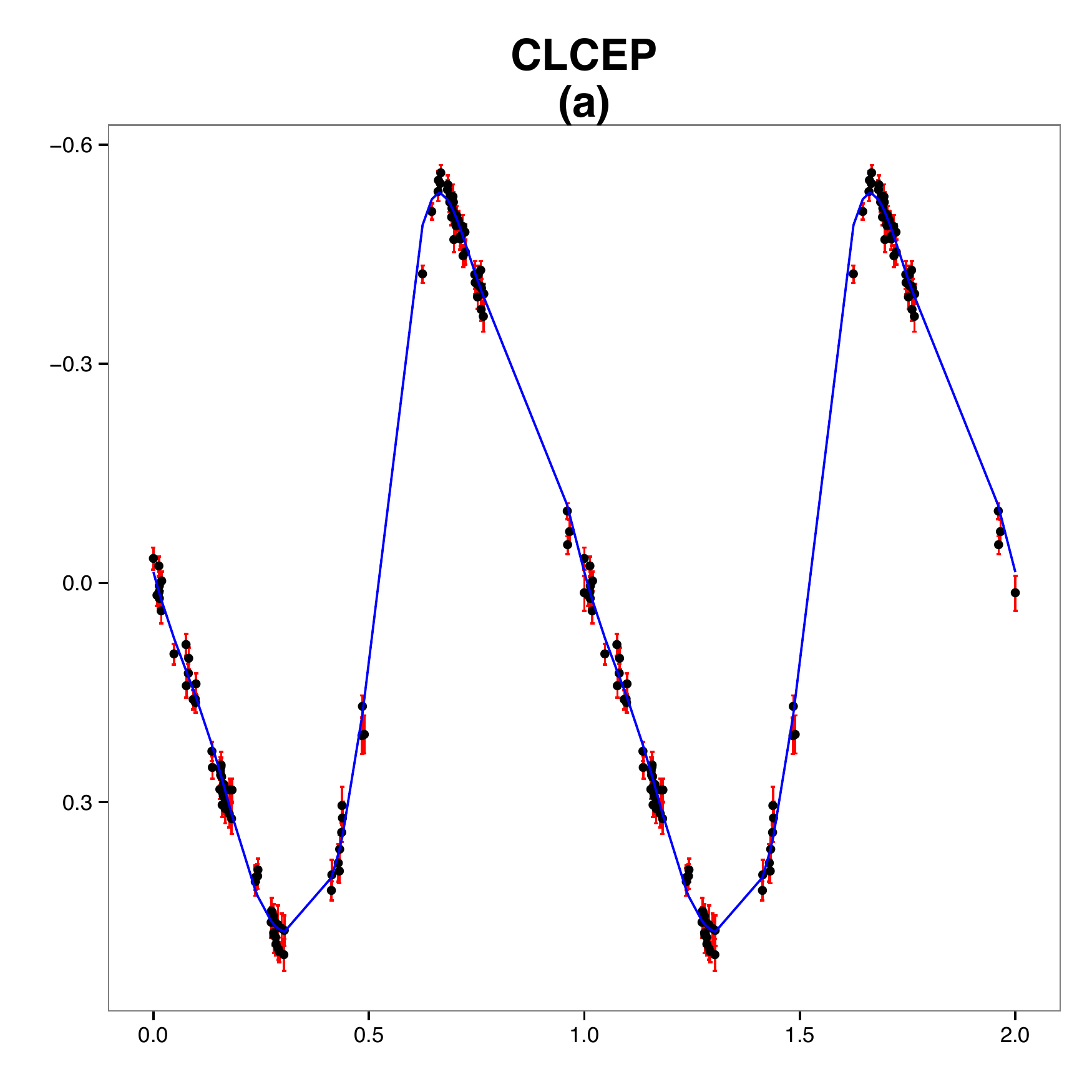}
\end{minipage}
\begin{minipage}{0.32\linewidth}
\includegraphics[width=\textwidth]{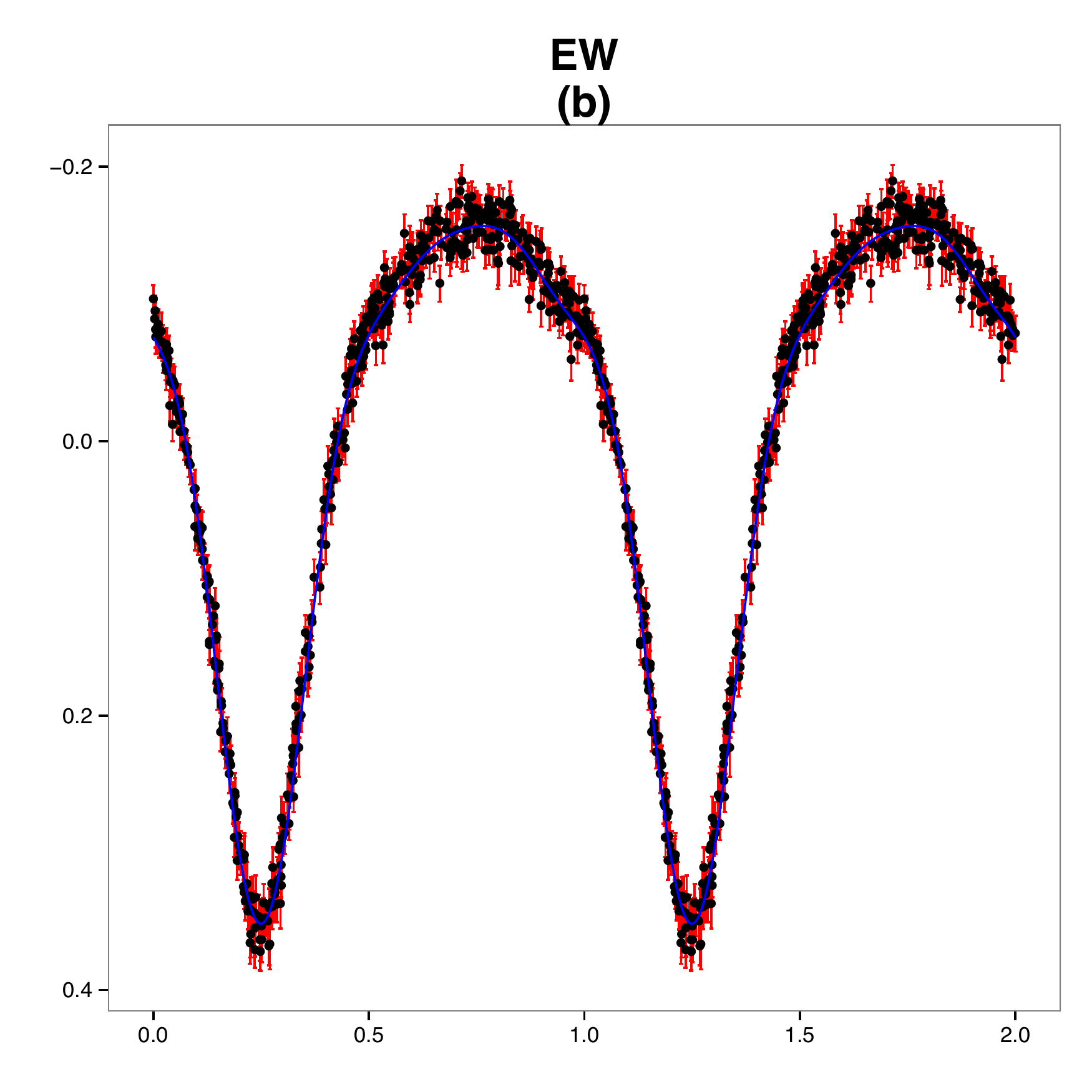}
\end{minipage}
\begin{minipage}{0.32\linewidth}
\includegraphics[width=\textwidth]{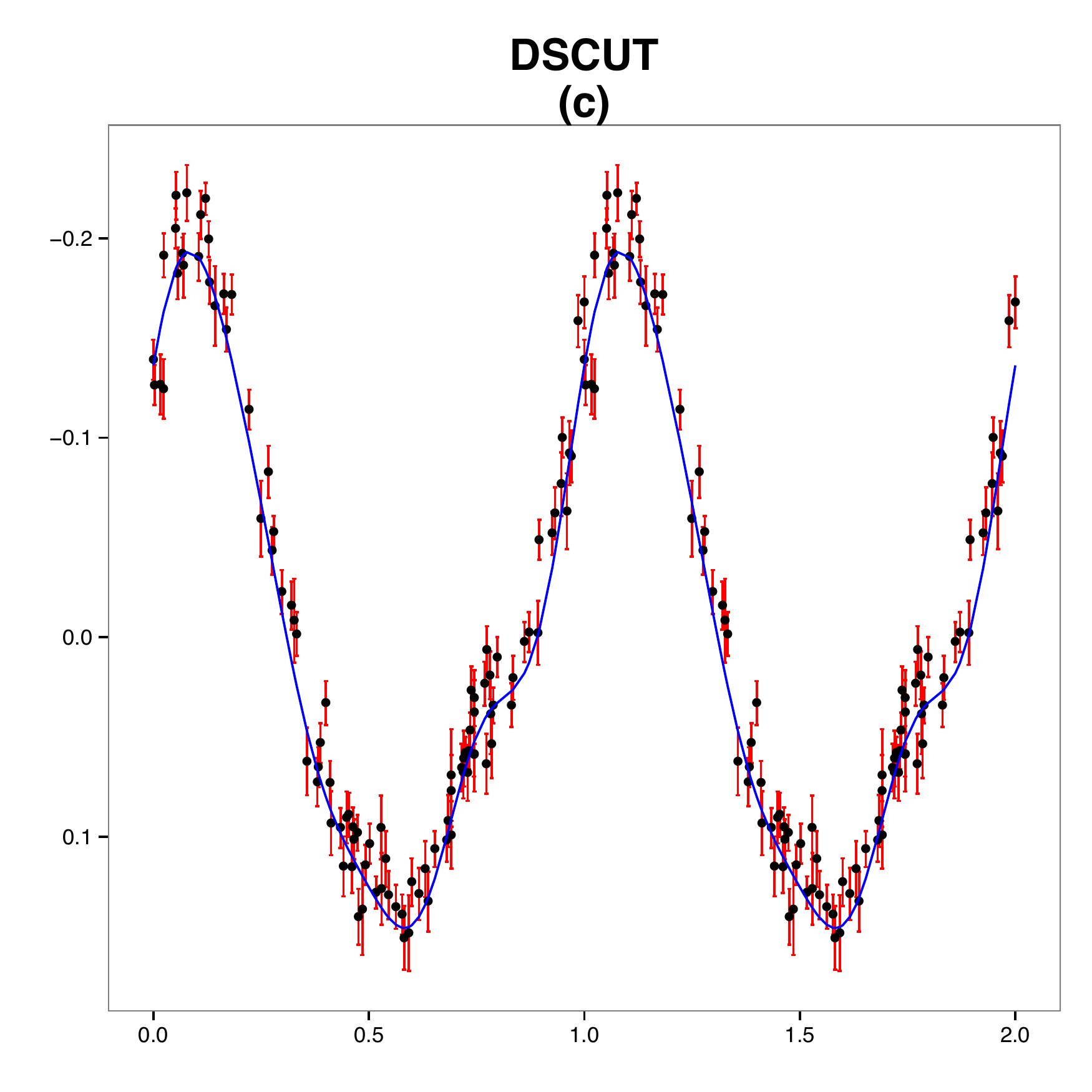}
\end{minipage}
\begin{minipage}{0.32\linewidth}
\includegraphics[width=\textwidth]{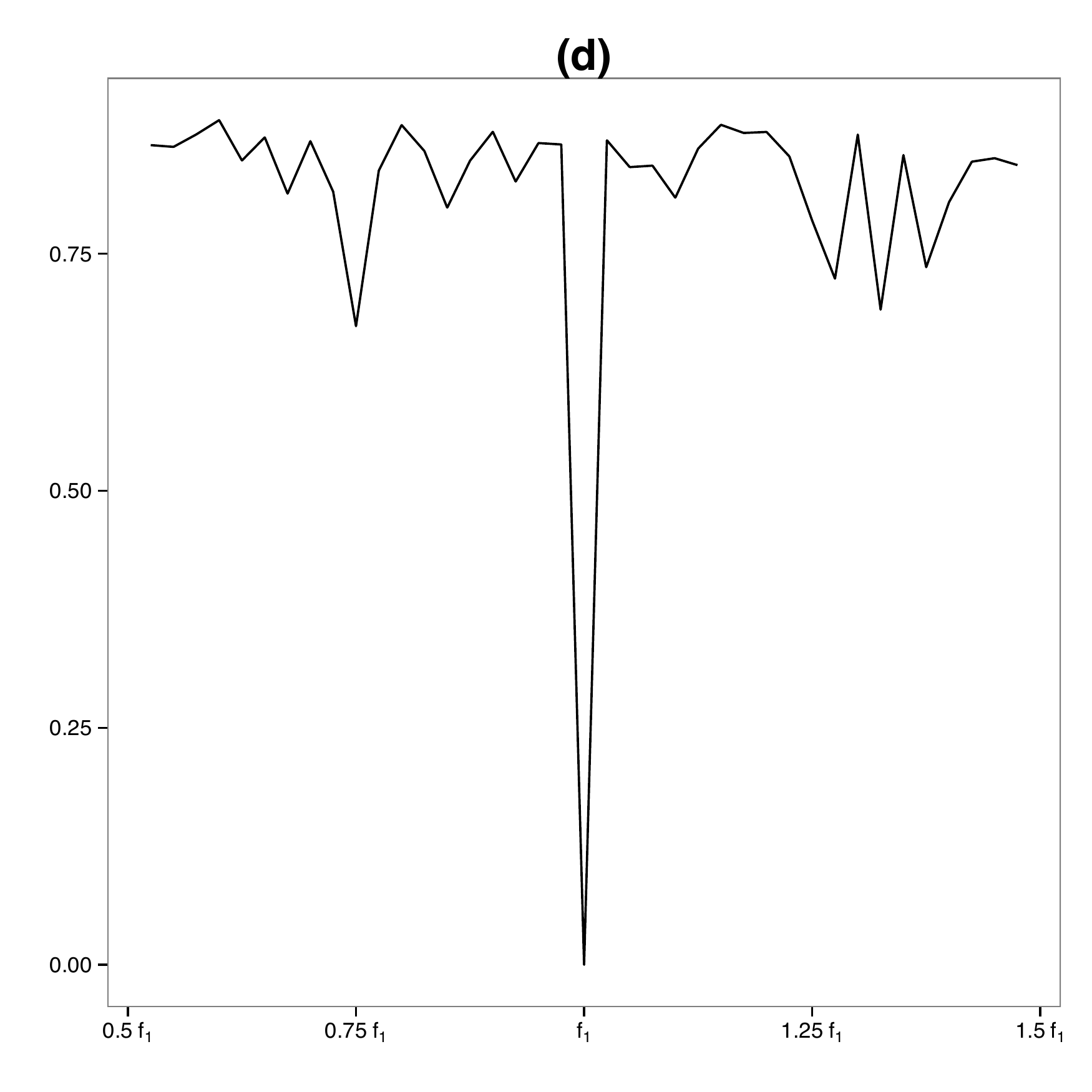}
\end{minipage}
\begin{minipage}{0.32\linewidth}
\includegraphics[width=\textwidth]{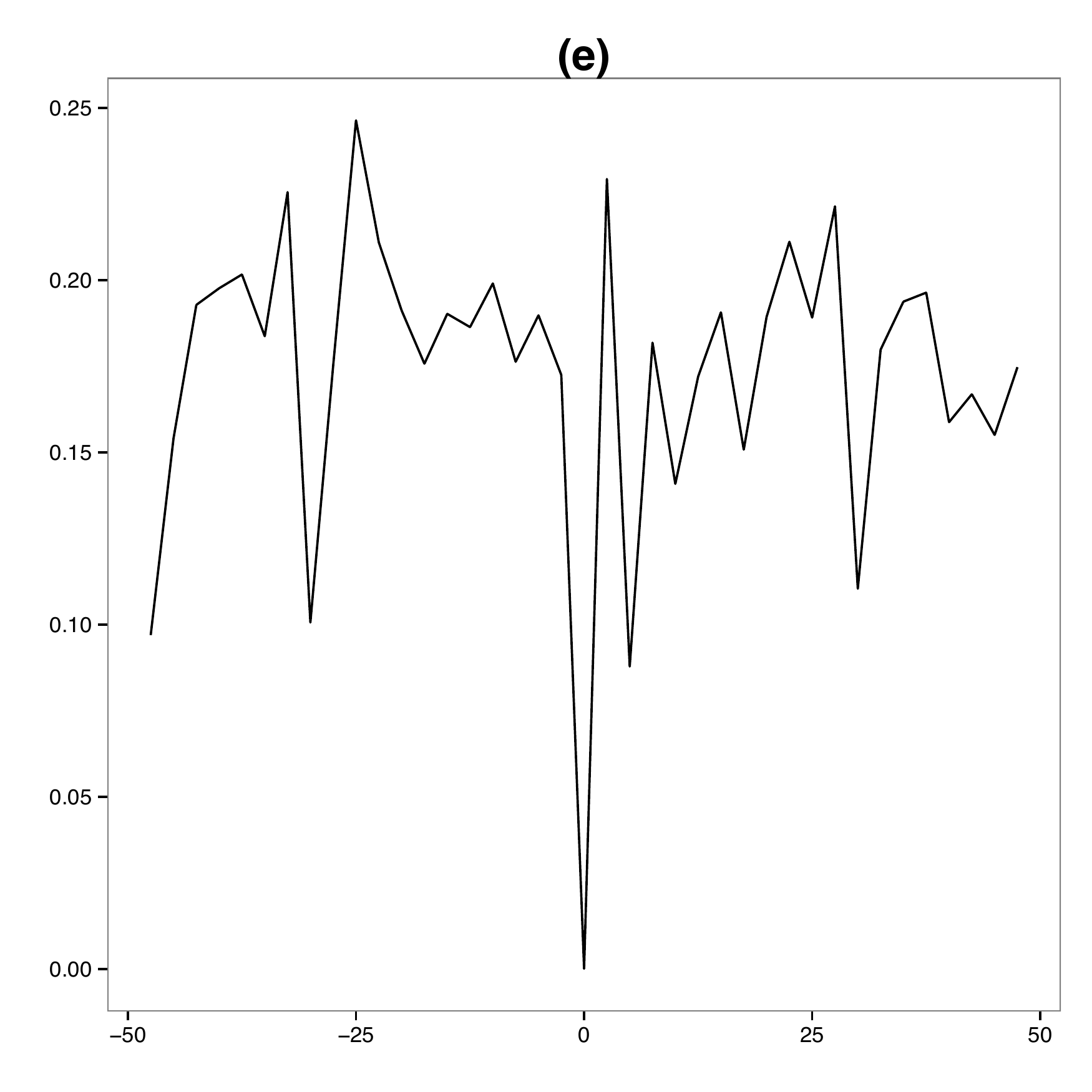}
\end{minipage}
\begin{minipage}{0.32\linewidth}
\includegraphics[width=\textwidth]{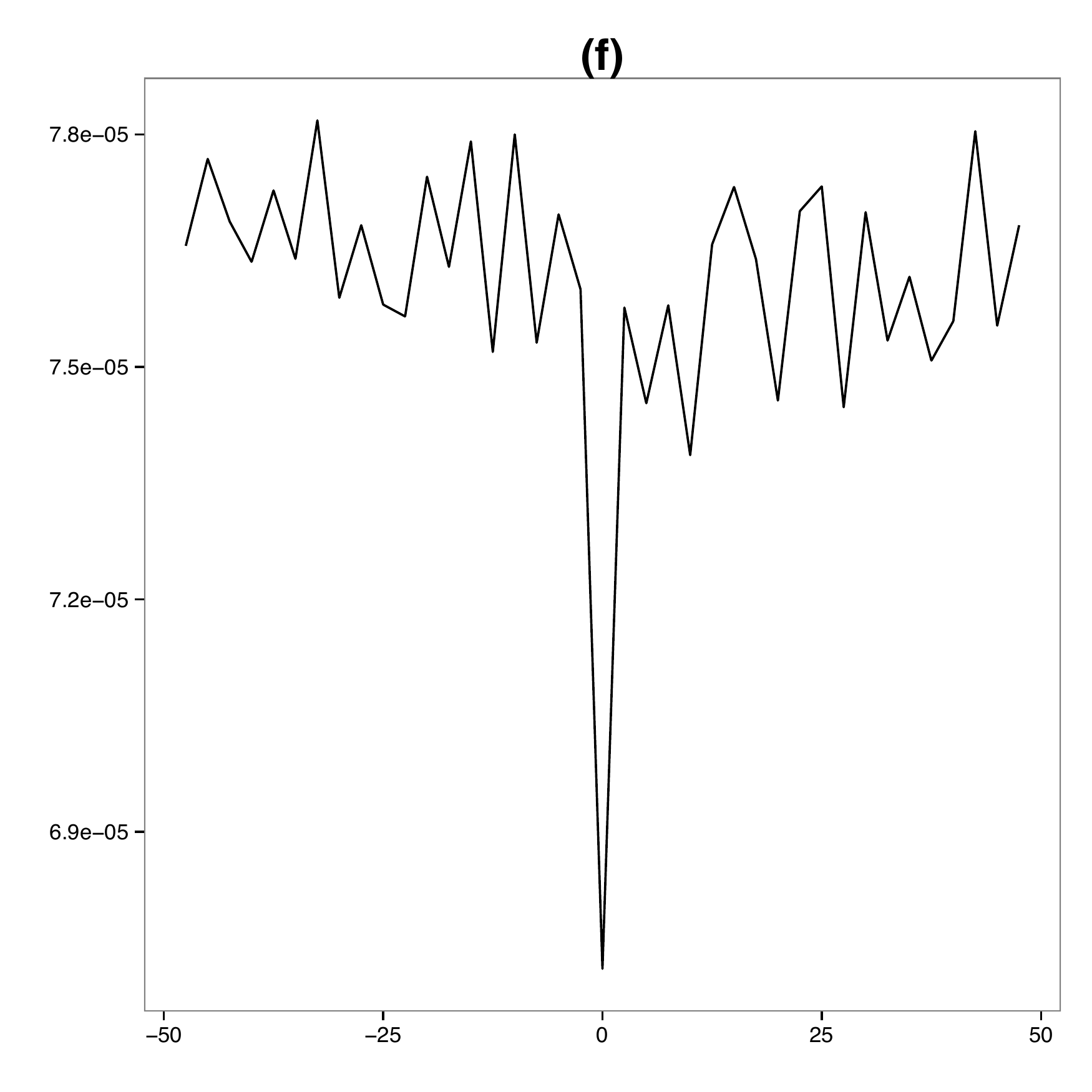}
\end{minipage}
\begin{minipage}{0.33\linewidth}
\includegraphics[width=\textwidth]{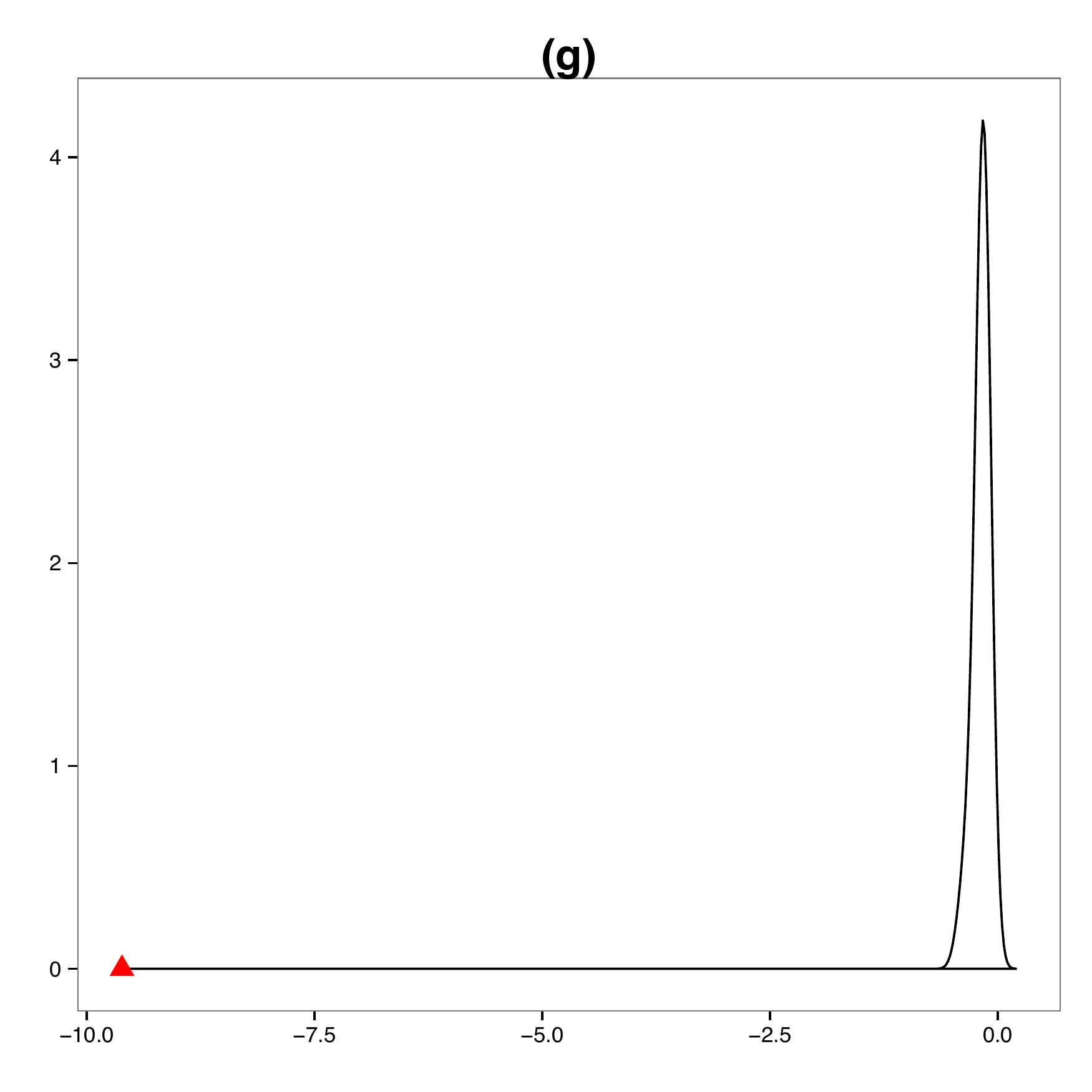}
\end{minipage}
\begin{minipage}{0.32\linewidth}
\includegraphics[width=\textwidth]{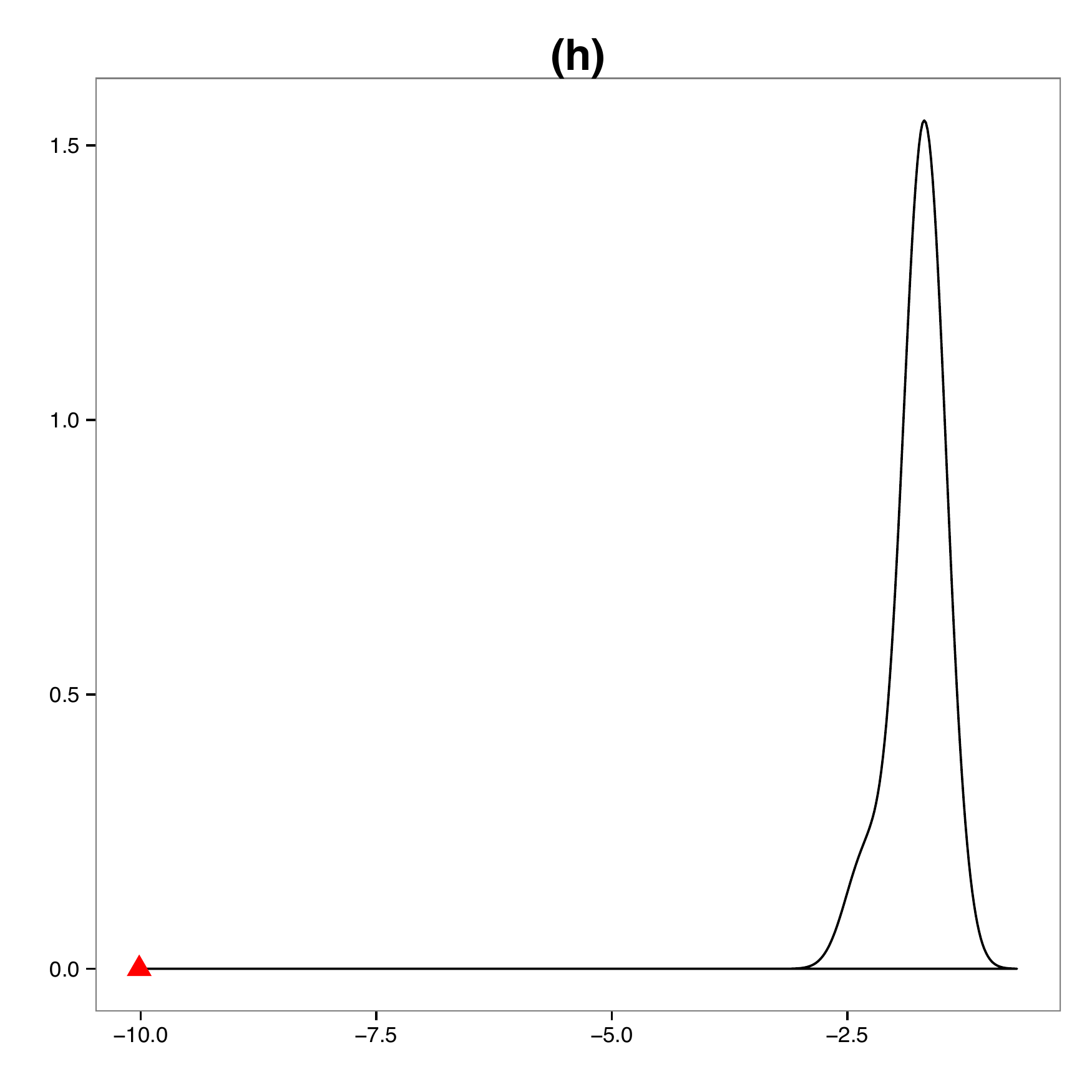}
\end{minipage}
\begin{minipage}{0.33\linewidth}
\includegraphics[width=\textwidth]{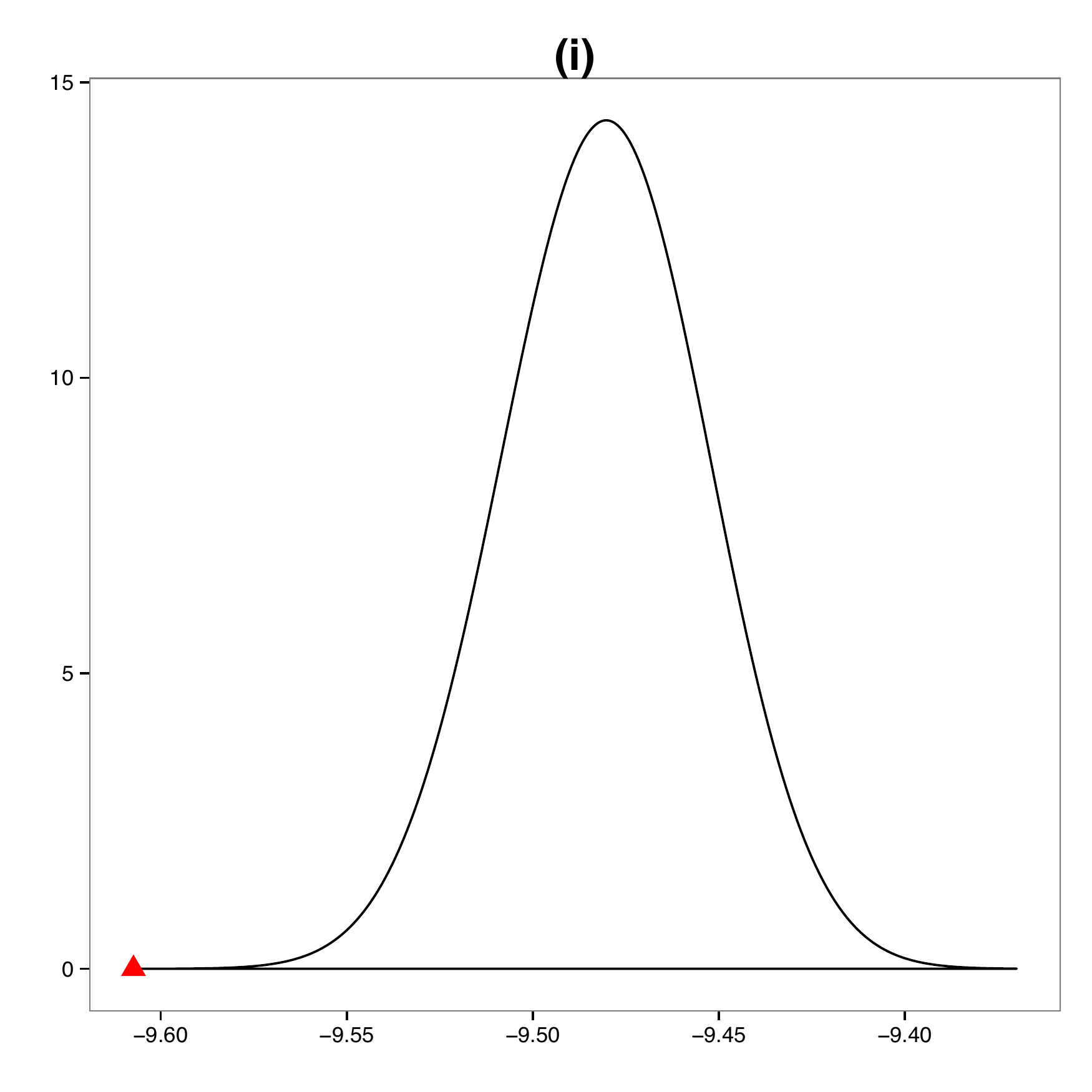}
\end{minipage}
\caption{In the first row,  the light curves of a Classical Cepheid, EW and DSCUT are shown on figures (a)-(c) respectively. The continuous blue line is the harmonic best fit. On the second row (figures (d)-(f)), for each of the variable stars, it is depicted on the x-axis the \% of variation from the correct frequency, and on the y-axis is the estimate of the parameter $\phi$ of the IAR model obtained after fitting an harmonic model with the wrong period (except at zero that corresponds to the right period). On the third row (figures (g)-(i)), the distribution of  the parameter $\phi$ of the IAR model is shown when each light curves is fit with the wrong period. The red triangle corresponds to the value of $\phi$ when the correct period is used in the harmonic model fitting the light curves.\label{fig:ex}}
\end{figure*}
\end{center}

First, we show that the model can be used to identify wrongly estimated periods. We select forty variable stars from the OGLE and Hipparcos surveys for which the harmonic model gives a precise fit of the light-curve. In selecting these stars we can be certain that the periods are well estimated.  These variable stars are selected from a group of $250$ stars which have the highest $R^2$ values in the harmonic fit. The multiple correlation coefficient, $R^2$, is a standard statistical measure for assessing goodness-of-fit.  In order to  take a representative sample of the classes and frequencies values observed in OGLE and HIPPARCOS, we binned the frequencies in five groups, and select eight light-curves from each bin and try at the same time to keep the representation of the classes of the original dataset. Figure~\ref{fig:ex}(a)-(c) show three examples of such set of light curves and Table~\ref{table:DistClass} the distribution of classes over the different frequency bins.

\begin{table*}
\centering
\begin{tabular}{|l|c|c|c|c|c|}
  \hline
Class  & $f_1 \leq 0.1$ & $0.1 < f_1 \leq 0.5$ & $0.5 < f_1 \leq 1$ & $1 < f_1 \leq 2$ & $f_1 > 2$  \\ 
  \hline
Classical Cepheid  (CLCEP) & 2 & 4 & & &\\ 
Chem. Peculiar (CP) &  & 1 & & &\\ 
Double Mode Cepheid  (DMCEP) & & & 1 & 2 & \\ 
Delta Scuti (DSCUT) & & & & & 2 \\ 
Beta Persei (EA) & & 1 & 4 & 2 & \\ 
Beta Lyrae (EB) & 1 &  & 2 & 2 & \\ 
W Ursae Maj (EW) & & 1 & 1 & 1 & 2\\ 
Mira  (MIRA) & 4 & & & & \\ 
PV Supergiants  (PVSG) & & 1 & & &  \\ 
RR Lyrae, FM (RRAB) & & & & 1 & 1 \\ 
RR Lyrae, FO (RRC) & & & & & 2 \\ 
Semireg PV (SR) & 1 & & & & \\ 
SX Phoenicis (SXPHE) & & & & & 1\\ 
   \hline
 Total & 8 & 8 & 8 & 8 & 8 \\ 
   \hline
\end{tabular}
\caption{Distribution of the forty examples selected by its frequency range and class of variable stars.}
\label{table:DistClass}
\end{table*}

\begin{figure}
\centering
\includegraphics[width=0.5\textwidth]{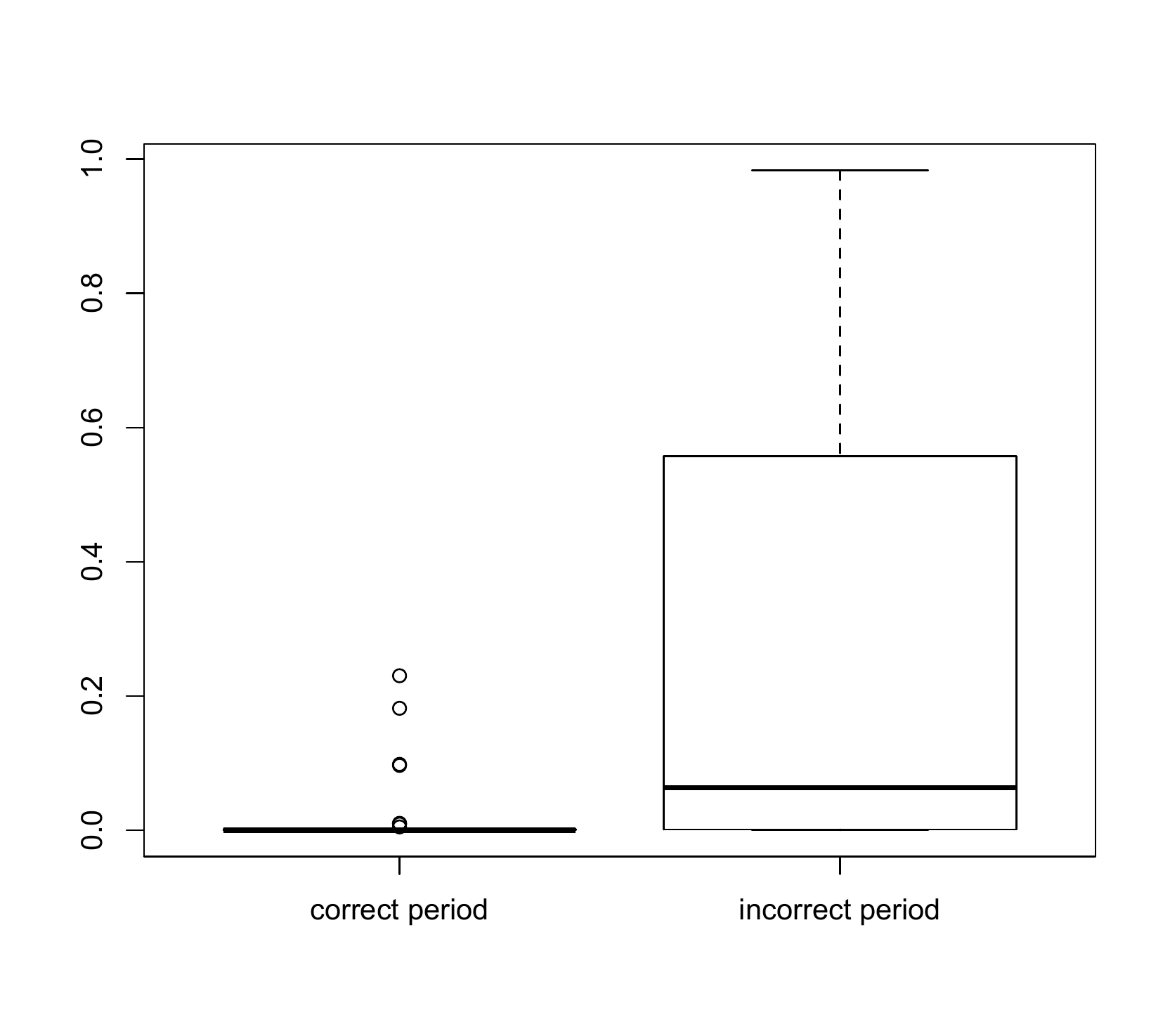}
\caption{Boxplot of the distribution of $\phi$, for the light-curves using the correct frequency (on the left) and for the light-curves using the incorrect frequency (on the right).\label{fig:boxplot}}
\end{figure}

We  apply the IAR model to the residuals of the best harmonic model, shown in equation (\ref{eq:harmonic}). For the forty chosen light-curves we obtain small values close to zero for the parameter $\phi$, as shown in the boxplot on the left of Figure~\ref{fig:boxplot}. This is  expected given that the model fits the light curves very well and thus the residuals are consistent with white noise. We then vary the frequency in the interval $(f_1-0.5f_1,f_1+0.5f_1)$ taking a total of $38$ frequencies equally space $g_1,\ldots,g_{38}$, $19$ to the right of the correct frequency $f_1$ and $19$ to the left. After doing so we fit the harmonic model with each wrong frequency $g_j$ taken from the interval. The residuals of the harmonic model have now temporal structure that can be captured with the IAR model, and in particular by the inferred value of $\phi$. For each ``incorrect" frequency $g_j$ we obtain a $\hat{\phi}_j$. 
The second row of Figure~\ref{fig:ex} shows the plot of the pairs $(g_j,\hat{\phi}_j)$  (with the right frequency $f_1$ at the center of the plot). Note that as we move away from the correct frequency, the value of $\phi$ generally increases in a non-monotonic way. Figure~\ref{fig:boxplot} shows in the  boxplot on the right the distribution of $\hat{\phi}$ for the light-curves with the incorrect frequency. This distribution is spread-out, taking in general large values away from zero, which reflects the correlation structure that remains.

Summarizing, for a given variable star and a period we fit an harmonic model and we then apply the IAR model to check whether there's any evidence of temporal structure which in this case would arise from period misspecification. If we obtain a $\hat{\phi} \neq 0$ we want to assess whether it is possible to conclude that there is significant temporal structure or not. In order to do that, we propose the following statistical test.

\subsubsection{Statistical test for assessing significance of the parameter $\phi$}

In the second row of Figure~\ref{fig:ex} we observe the relationship between frequency of the variable stars versus the parameter $\phi$ of the autoregressive model. At zero in the x-axis lies the correct frequency for which we obtain the smaller $\phi$ value in the three examples shown. This is expected because the light-curves are chosen such that the harmonic model attains an accurate fit. Note that even though the smaller $\phi$ is obtained at the estimated frequency $f_1$, this value of $\phi$ relative to the neighbouring values differ substantially. Note also that while the graph  on the left has values of $\phi$ above $0.75$, in the middle the values are around $0.18$  and in the figure on the right all values are  between $7.5\times 10^{-5}$ and $7.8\times 10^{-5}$, with the exception of the value of $\hat{\phi}$ at its minimum in $f_1$ which as expected is close to zero. Therefore, just from the value of $\hat{\phi}$ it is not always possible to discriminate between a correct period with residuals without temporal dependency and an incorrect period with residuals with temporal dependency. We propose to evaluate whether the minimum $\hat{\phi}$ is significantly smaller than the remaining $\hat{\phi}$ by assuming that the log$(\hat{\phi})$ distributes as a Gaussian. The bottom row of Figure~\ref{fig:ex} shows the density of the log($\hat{\phi}$) values at the incorrect periods, and the red triangle shows the log($\hat{\phi}$) values at the correct period. 
The $p$-values for the three log($\hat{\phi}$) at the correct period are  $0, 1.62\times 10^{-281}, 2.86\times 10^{-19}$ respectively, indicating that they are all statistically significantly smaller that their neighbours.

\subsection{Study on simulated and real multiperiodic variable stars}

Several classes of variable stars can have multi-periodic stars, for example, double-mode Cepheids and double-mode RR-Lyrae. For those stars fitting an harmonic model with only one period produces errors in the model that are not independent, but correlated. Therefore, we expect that when fitting the  IAR model to the  residuals of this harmonic model, the estimate of the parameter $\phi$ will be large. We show with simulated and real data that this is indeed the case, illustrating a case where the model lacks the complexity to describe the time series at hand.

\begin{center}
\begin{figure*}
\begin{minipage}{0.45\linewidth}
\includegraphics[width=\textwidth]{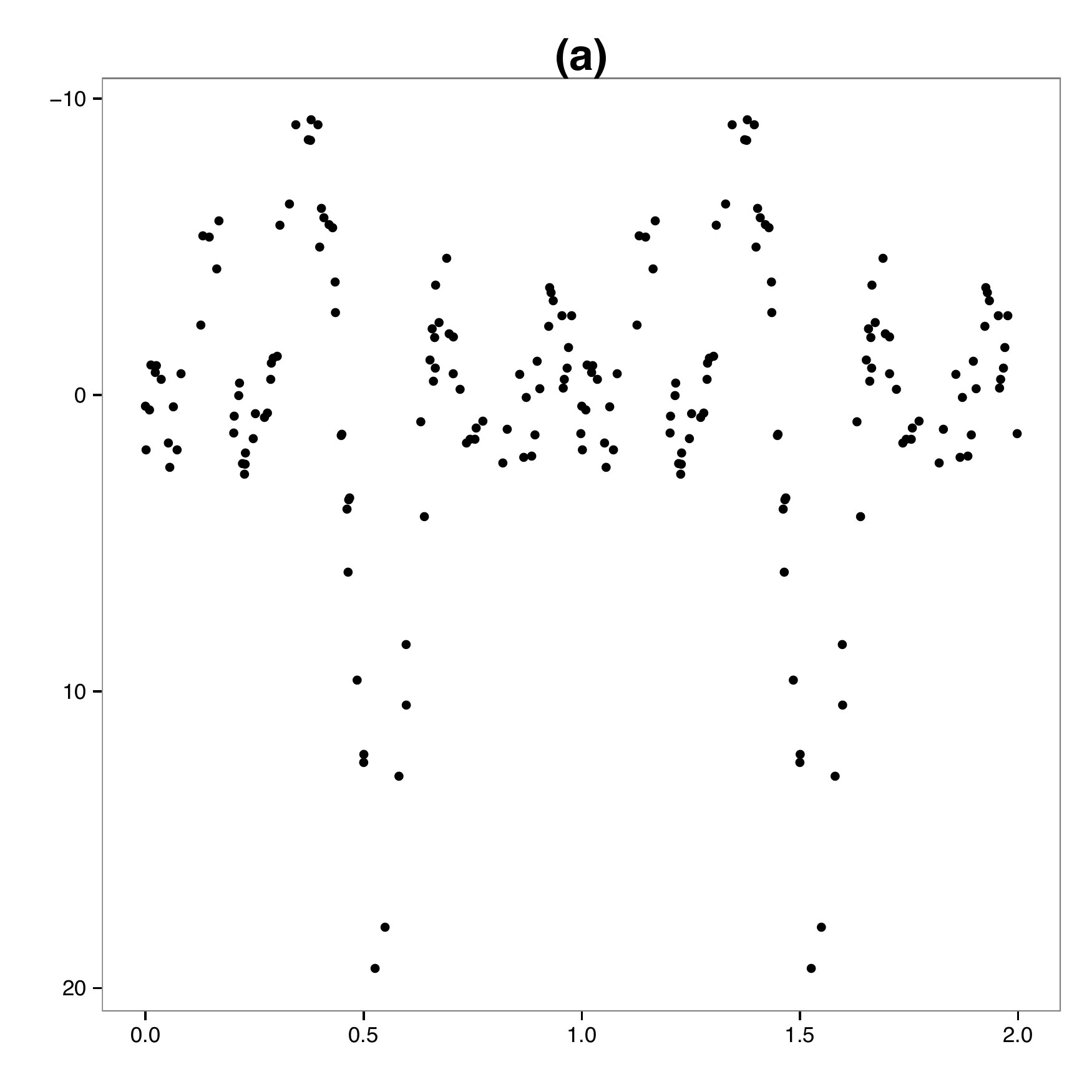}
\end{minipage}
\begin{minipage}{0.45\linewidth}
\includegraphics[width=\textwidth]{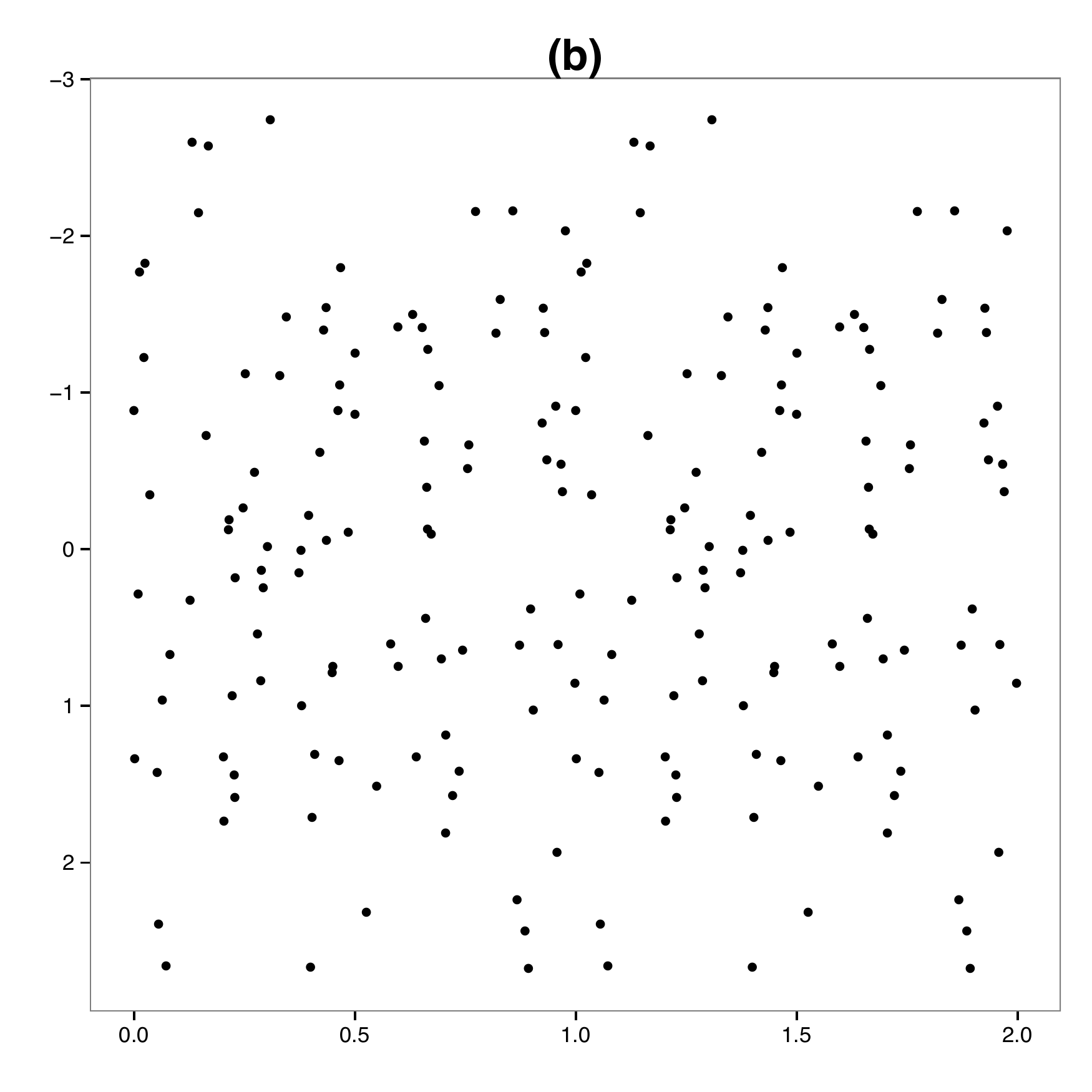}
\end{minipage}
\caption{(a) Residuals of the best harmonic  fit with one frequency for a simulated multiperiodic light curve; (b) Residuals of the best harmonic best fit with two frequencies  for the same simulated multiperiodic light curve.\label{fig:multiperiodicSintetic2}}
\end{figure*}
\end{center}

We simulate  multi-periodic light-curves with two periods using the harmonic model. We show an example in which the light curve is simulated using the harmonic model with two periods and four components for each period. Specifically, at time $t$ the value simulated is $y(t)=\mathop{\sum}\limits_{i=1}^2\mathop{\sum}\limits_{j=1}^4 ( sin(2\pi f_i jt) + cos(2\pi f_i jt)) +\tau_t$, where $\tau_t$ is generated from a standard Gaussian distribution with mean zero and variance one and $f_1=1/3, f_2=1/12$. The observational times are simulated using a mixture of two exponential distributions, i.e. $f(t|\lambda_1,\lambda_2,w_1,w_2)=w_1g(t|\lambda_1)+w_2g(t|\lambda_2)$, where $\lambda_1=130$ and $\lambda_2=6.5,w_1=0.15,w_2=0.85$.In Figure \ref{fig:multiperiodicSintetic2} we show on the top plot the residuals after fitting an harmonic model with one period. The header has the value of $\phi =0.5447$, which is the value of $\phi$ estimated from these residuals. The bottom plot has the residuals after fitting  the harmonic model with two periods. From this series the estimated of $\phi$ has decreased to a small value close to zero  ($\leq 0.0001$).  

\begin{center}
\begin{figure*}
\centering
\begin{minipage}{0.45\linewidth}
\includegraphics[width=\textwidth]{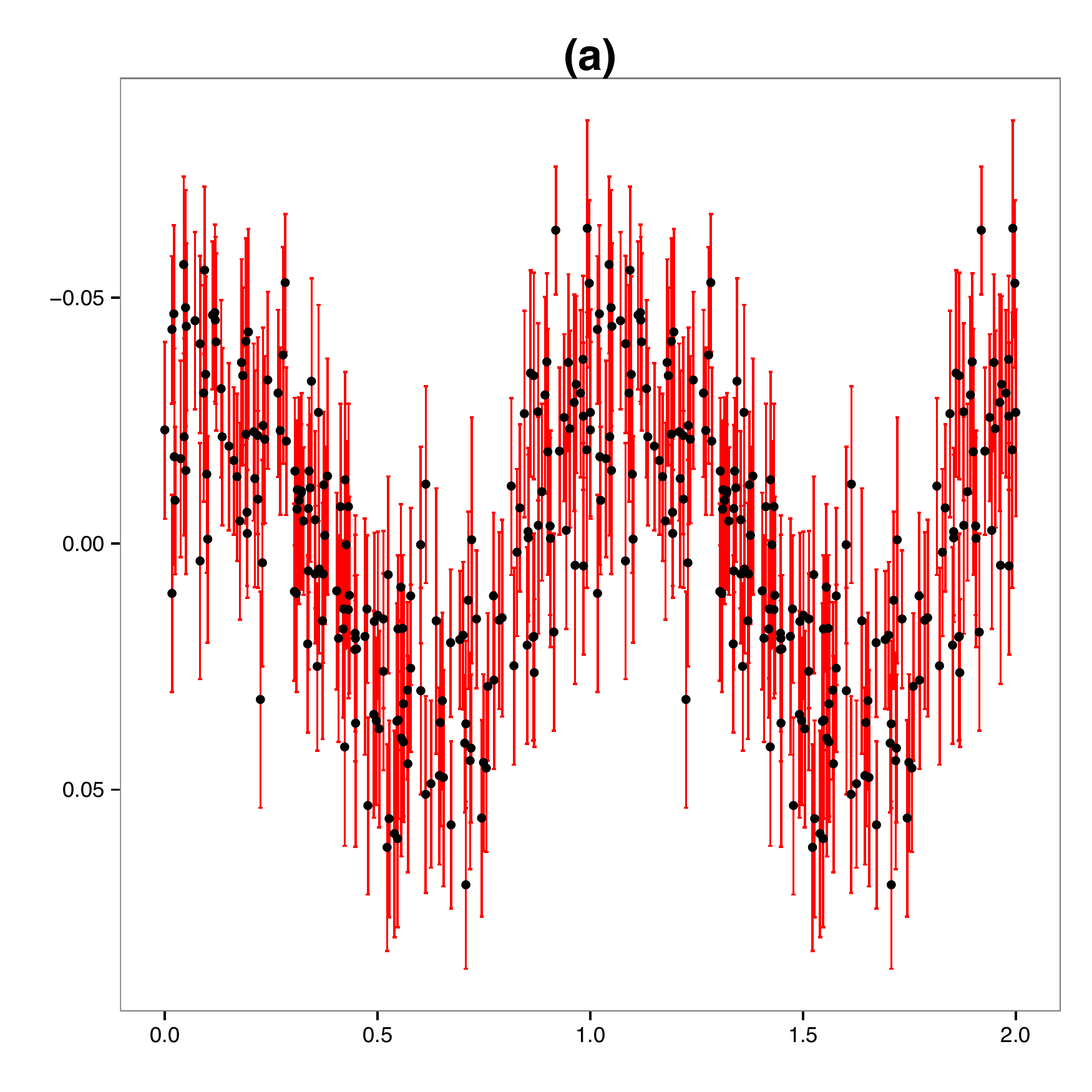}
\end{minipage}
\begin{minipage}{0.45\linewidth}
\includegraphics[width=\textwidth]{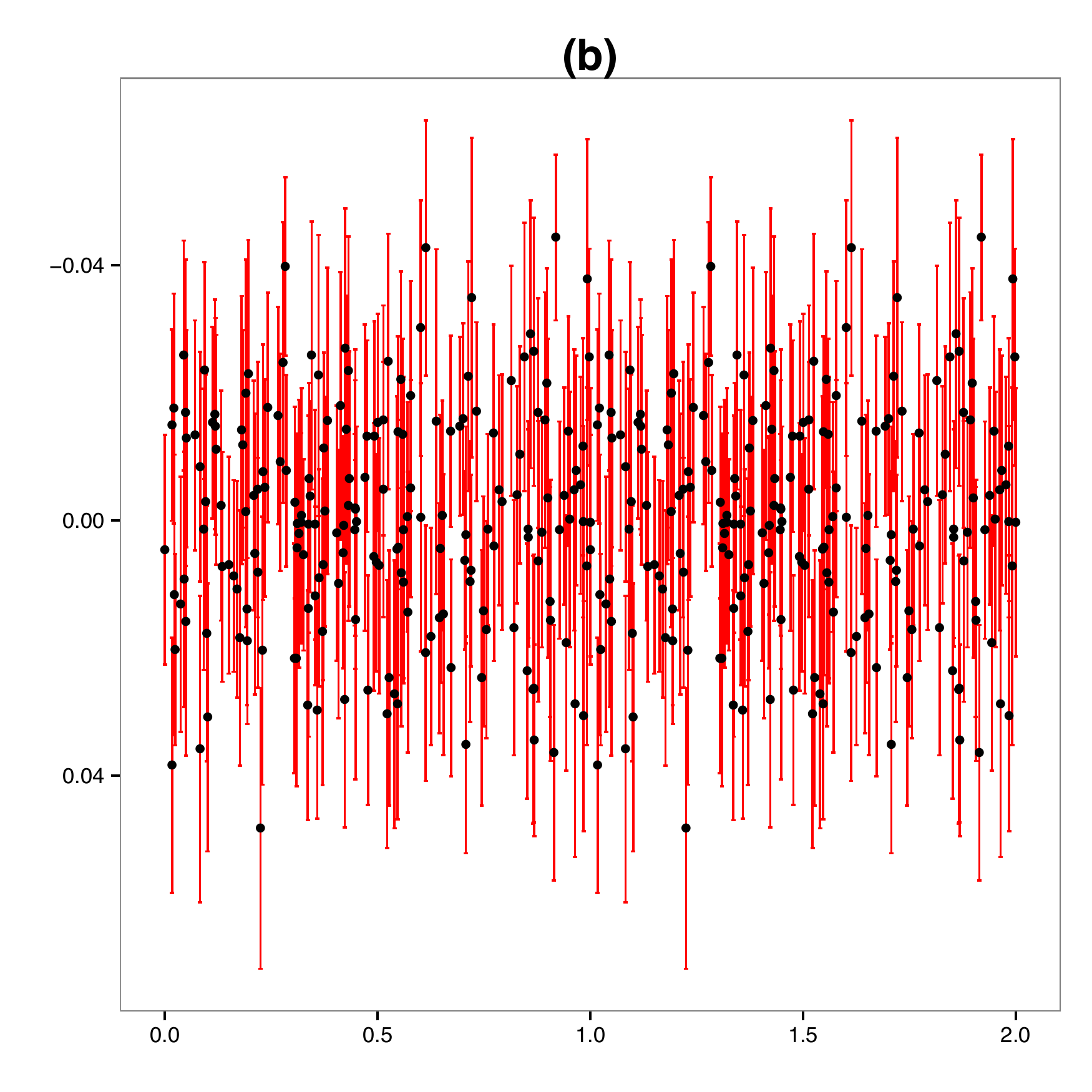}
\end{minipage}
\caption{a)  Residuals of a double model Cepheid after fitting an harmonic model with one period; b) residuals of the same variable star after fitting an harmonic model with two periods.}
\label{fig:multi}
\end{figure*}
\end{center}

From the set of real light curves observed in the OGLE and Hipparcos surveys we identified some multiperiodic variable stars. Figure~\ref{fig:multi}a)  shows the residuals of a double model Cepheid after fitting an harmonic model with one period and  Figure \ref{fig:multi}b) shows the residuals of the same variable star after fitting an harmonic model with two periods. The $\hat{\phi}$ of the IAR model at the residuals after fitting an harmonic model with one period is $0.5411$ while $\hat{\phi}=0.033$ at the residuals after fitting an harmonic model with two periods.

\subsection{Exoplanet Transit light-curve}
\label{sec:planet}

\begin{figure}
\centering
\includegraphics[width=0.5\textwidth]{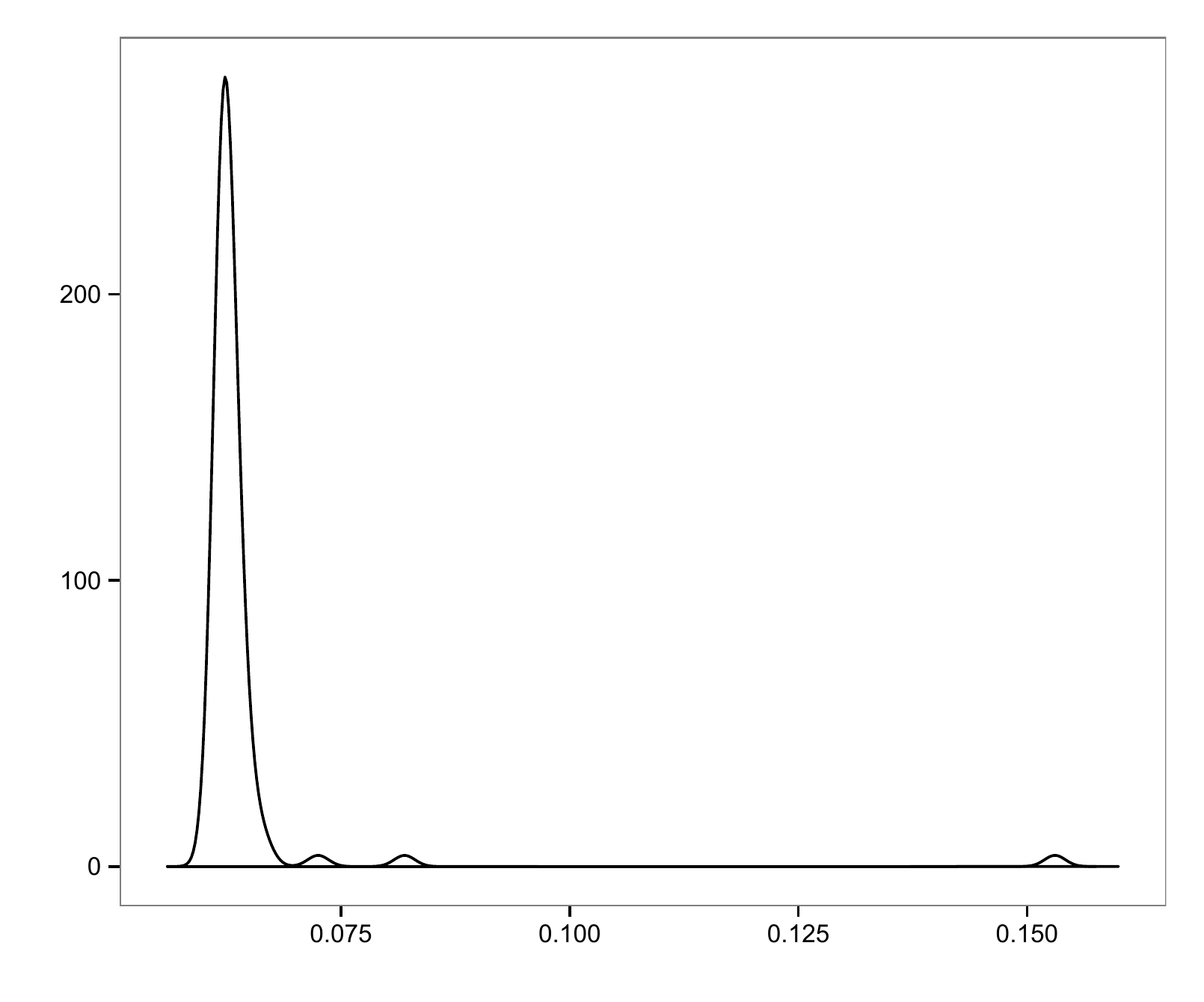}
\caption{Density of the observational time gaps from WASP-6\label{fig:res}}
\end{figure}

A planet orbiting a star will block part of the signal if it transits in front of it as seen from our vantage point. The  observed flux can then be modelled by multiplying the approximately constant flux of the star with the transit signal, which can be modelled with the formalism described in \cite{Mandel_etal02}. We have again a structure for the model described by $z(t)= g(t,\theta)+\epsilon(t)$, where $z(t)$ represents in this case the logarithm of the measurement flux of the star, $g(t,\theta)$ is the sum of a log constant flux and the transiting signal and $\epsilon (t)$ is the error at time $t$ assumed to be independent Gaussian with mean zero and variance $\sigma^2$. It is common that the residuals are not well modelled by white noise, and this can lead to biases in the estimation of transit parameters and their uncertainties e.g. \cite{carter:2009}. In \cite{Jordan_etal13} a transit of the exoplanet WASP-6b was observed with Magellan in order to estimate its transmission spectra. The white-light curve (time series of the stellar flux integrated over wavelength) was fit with a transit model and via a model-comparison process it was assessed that the residual structure was best described by a flicker model with power spectral density $\propto 1/f$,  indicating a long memory time dependency. Other models tried where a white noise model and an ARMA(2,2) model. All models tried assumed that the observational times are equally spaced, which in their case is a good approximation to the data, although it is not exact. In Figure~\ref{fig:res} we show some statistics of the time gaps between the observations for this data, which we will use to illustrate the performance of our model  on a dataset which should be very well suited for methods that assume constant cadence but that does have some small departures from such behaviour.

\begin{center}
\begin{figure*}
\begin{minipage}{0.48\linewidth}
\includegraphics[width=\linewidth]{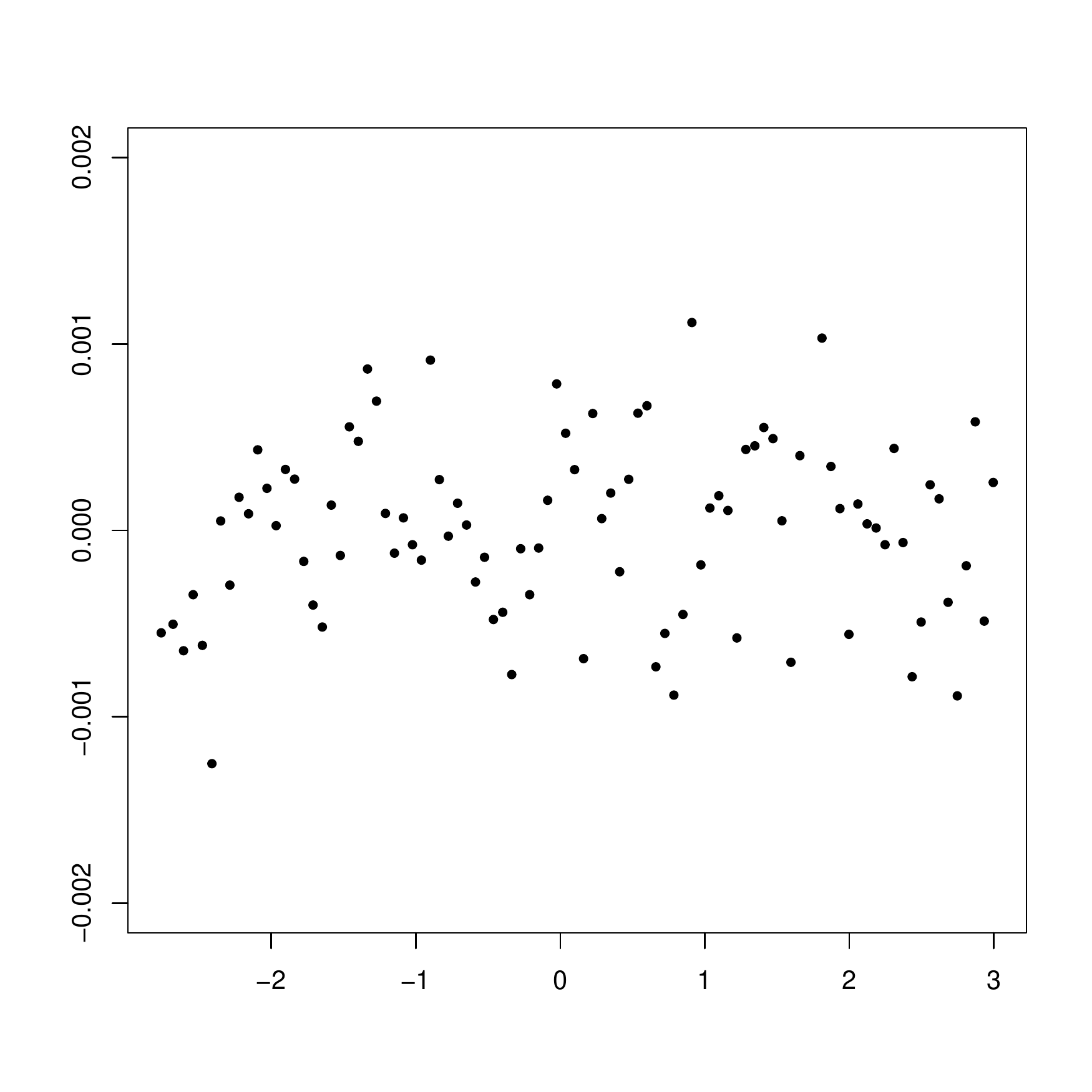}
\end{minipage}
\begin{minipage}{0.48\linewidth}
\includegraphics[width=\linewidth]{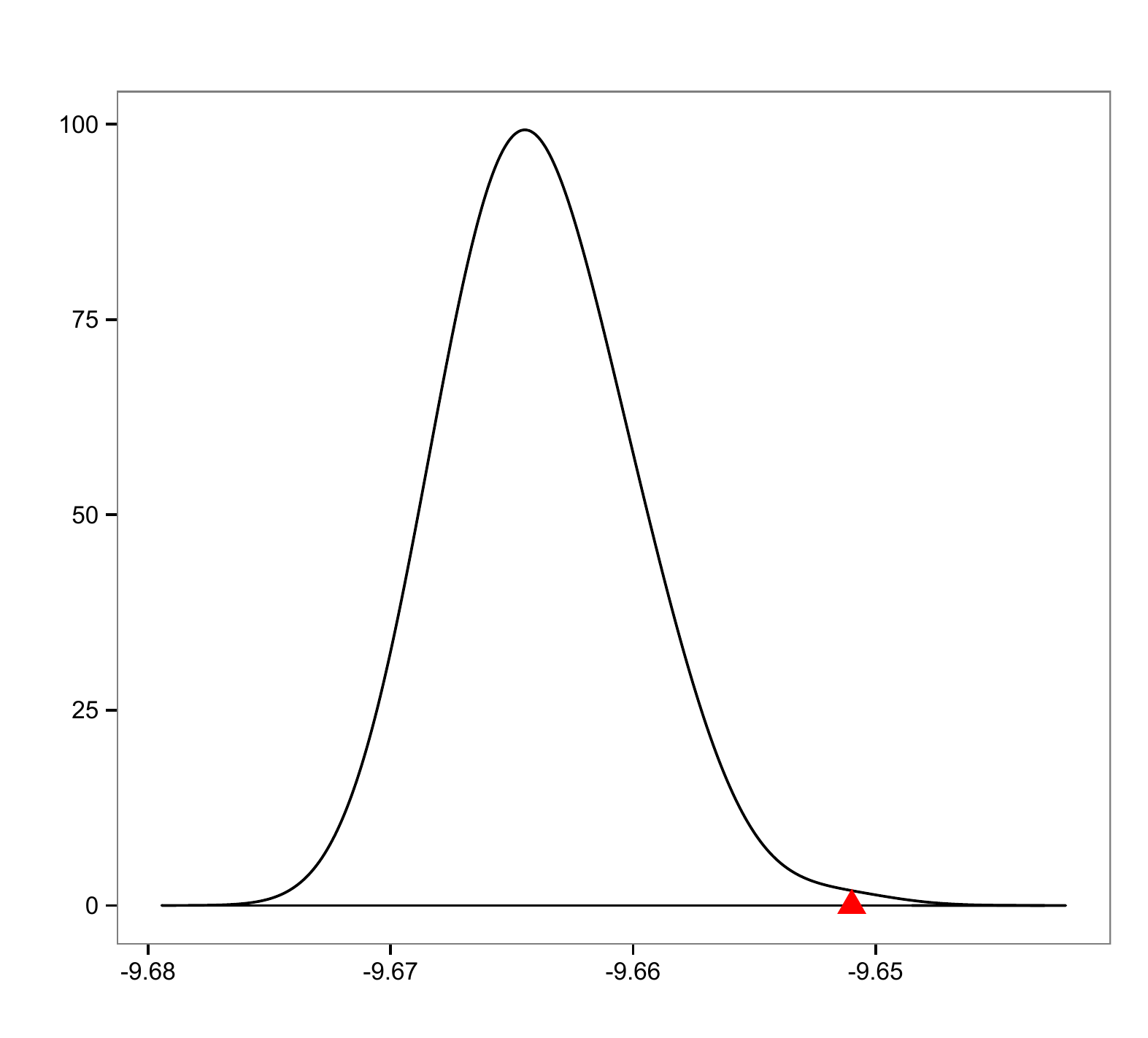}
\end{minipage}
\caption{(a) Residuals after fitting the model for a transiting exoplanet; (b) The red triangle represent the log($\hat{\phi}$), where $\hat{\phi}$ is the parameter of the IAR model. The black line represents the density of the $\phi$ for the randomized experiment.\label{fig:planet}}. 
\end{figure*}
\end{center}

After fitting the model, described above, to a star with an exoplanet orbiting around it, we implement the IAR model on the residuals, which are shown in Figure~\ref{fig:planet}(a). These residuals correspond to the same data utilized at  \cite{Jordan_etal13} and are shown in the left-bottom panel of Figure~6 in \cite{Jordan_etal13}.  The red triangle in Figure~\ref{fig:planet}(b) corresponds to log($\hat{\phi}$), where $\hat{\phi}$ is the estimator of the parameter of the IAR model. To evaluate whether this value of the parameter could have been obtained from a series with no temporal dependency, we perform a randomized experiment. In this experiment we fixed the observation times of the time series, but shuffled the flux measurements a hundred  times to obtain hundred estimates of the parameter $\phi$, which allow us to have an estimate of the $\phi$ values that are expected to be observed when there is no temporal dependency in the time series. This distribution is shown in Figure~\ref{fig:planet}(b). Note that the actual value of $\hat{\phi}$ is very unlikely to have arisen from this distribution, having a p-value of $5.64\times 10^{-5}$. This result is consistent with the results of \cite{Jordan_etal13}, where they also find temporal structure on this data using a flicker-noise and an ARMA model.

\noindent \textbf{Remark 1.} Observe that  the {\em residuals} of the fitted model, defined as $y_t=z_t-g(t,\widehat{\theta})$, are not necessarily equal to the model {\em errors} $\delta_t$, say. However, under the assumption that the estimator of $g(t,\theta)$, $g(t,\widehat{\theta})$ is consistent we have that asymptotically, $y_t \sim \delta_t$.  Note that due to the irregularity of the observation times, the residuals do not share the same variance. A well known procedure for assessing that the residuals are indeed white noise is the Ljung-Box test. Thus, we suggest to apply first this test to the adequately standarized residuals for whiteness, taking in consideration the sensibility of this test to the sample size. If the null hypothesis of white noise is rejected, then proceed to model the serial dependence observed in the residuals. Notice that the ultimate goal of this modeling approach is to obtain white noise residuals, that is, to remove all systematic error components. In the normal case,  the theoretical residuals are correlated when the covariates are not orthogonal, which is standard in multiple linear regression.  But in this case there are statistical tests, such as the Durbin-Watson type of test or Breusch-Godfrey test or Ljung-Box test, that  assess whether the  residuals remain correlated/autocorrelated. These tests have been extensively used in multiple linear regression. The purpose of the IAR model is  to test whether there remain significant correlation on $y_t$ and to model it.

\section{Discussion}
\label{sec:discussion}

In this work we present an autoregressive model for irregularly observed time series (IAR), and we show that it is weakly stationary, and under some conditions, it is stationary and ergodic, providing a solid statistical framework to assess autocorrelation in the residuals of a model sampled at irregular times. We show that this model is not limited by Gaussian time series. We develop  examples with samples from a Gamma and a Student-t distributed series, in which the IAR model under the correct distribution outperforms the model under the Gaussian distribution. 
We further develop statistical tests to assess significance of the parameter of the model that measures autocorrelation of the time series. We have developed a maximum likelihood procedure to estimate the model and provide code in the R statistical software and in Python.

We have illustrated two implementations of the model on astronomical dataset to show some possible applications in this field for the identification of misspecified models and the assessment of the presence of time correlated structure in time series. In both examples, we follow a two-stage approach for parameter estimation, i.e. first the parameter of the harmonic model are estimated and to the residuals of this model we implement and estimate the IAR model.  This is certainly not ideal, as it would be more appealing to jointly estimate the parameter of the IAR and the harmonic model. We have not presented it in that form because of the examples that we have chosen. Periodic light-curves from variable stars require to have a period estimated. While there are methodologies that estimate jointly the period and a parametric model (e.g. the coefficients of truncated Fourier series, \cite{Palmer_09}) by far the most common practice is to first estimate a period and then estimate the model parameters given a period (see, e.g., \cite{Elorrieta_etal16} and references therein). We follow the same procedure with the light-curve of the star with an orbiting exoplanet. For other implementations,  we advocate simultaneous estimation of the parametric and IAR models.

The model presented here is a simple model that depends on one parameter that measures the autocorrelation of the series and another parameter that measures the size of the error of the model. Nevertheless, having correlated errors not accounted for in the specification of a model can have important consequences. For example, in the context of linear regression, the estimator of the error of the model can be biased toward zero. This can lead to confidence intervals that are too narrow, based on the t-statistic, and therefore,  can produce falsely significant results.

A drawback that both the IAR model and the CAR(1) have, is that they only allow to estimate positive autocorrelation, i.e. the parameter $\phi$ is constraint to be non-negative. In the case of the Gaussian and non-Gaussian IAR models,  equation (\ref{Model}) would require a negative $\phi$ to the power of a real number which, in general, does not exist. 
In the case of the CAR(1) model, the autocorrelation is measured by $e^{-\alpha_0}$, where $\alpha_0>0$ for the process to be stationary, and therefore also takes only positive numbers. We are currently extending the IAR model to allow to estimate series  with negative autocorrelation. 

With this work we try to entice the researchers to model time series with irregular times as series of discrete and not continuous times as they have been commonly treated. This opens a new avenue for developing models that can fit irregular time series based on discrete times. These models can be simple but with sound statistical properties. We consider that the discrete representation for irregular time series is specially suitable for time series obtained from astronomical datasets because the gaps between observations can be very large, in the order of days, months or years. Whereas in disciplines where the time gaps between observations are tiny, a continuous model such as the continuous autoregressive model could be more suitable.

{\it Software to implement the model and simulations are available in Python and R upon request to the authors.}

\section*{Acknowledgements}

Support for this research was provided by grant IC120009, awarded to The Millennium Institute of Astrophysics, MAS, and from Fondecyt grant 1160861. F.E. acknowledges support from CONICYT-PCHA (Doctorado Nacional 2014- 21140566).

\bibliographystyle{mnras}
\bibstyle{mnras}
\bibliography{iar.bib}

\appendix

\section{Representation of CAR(1) model in a form of a discrete irregular time series}
\label{sec:car1}
 
As mentioned in section \ref{sec:car} the CAR(1) model is defined as the solution of the following stochastic differential equation

\begin{equation}
\frac{d}{dt}\epsilon (t)+\alpha_0 \epsilon (t)=\sigma_0 \nu (t) +\beta,
\end{equation}

\noindent
where $\nu (t)=\frac{d}{dt}w(t)$ and $w(t)$ is a Brownian motion or Wiener process. The derivative of $w(t)$ does not exist, so a proper way of writing equation (\ref{eq:car1}) is as an It$\hat{\mbox{o}}$ differential equation

\begin{equation}
\label{eq:cardif}
d \epsilon (t) +\alpha_0 \epsilon (t) dt = d w(t) +\beta dt,
\end{equation}

\noindent 
where $d \epsilon (t)$ and $d w(t)$ denote the increments in $\epsilon$ and $w$ in the time interval $(t,t+dt)$, and $\epsilon(0)$ is a random variable with finite variance and independent of $\{w(t)\}$. 

The solution of equation (\ref{eq:cardif}) can be written as

\begin{equation}
\label{eq:carsol}
d \epsilon (t)=e^{-\alpha_0 t}\epsilon (0) +e^{-\alpha_0 t} I(t) +\beta e^{-\alpha_0 t}\int_0^te^{\alpha_0 u}du,
\end{equation}

\noindent 
where $I(t)=\sigma_0\int_0^te^{\alpha_0u}dw(u)$ is an It$\hat{o}$ integral satisfying $E(I(t))=0$ and $\mbox{Cov}(I(t+h),I(t))=\sigma_0^2\int_0^te^{2\alpha_0u}du$ for all $t\geq 0$ and $h>0$. It can be shown that necessary and sufficient conditions for $\{\epsilon (t)\}$ to be stationary are $\alpha_0>0, E(\epsilon(0))=\beta/\alpha_0$ and $Var(\epsilon(0))=\sigma_0^2/(2\alpha_0)$. Further, if $\epsilon (0)\sim N(\beta/\alpha_0,\sigma_0^2/(2\alpha_0))$, then the CAR(1) process is also Gaussian and stationary.

If $\alpha_0>0$ and $0\leq s \leq t$, it follows from equation (\ref{eq:carsol}) that $\epsilon(t)$ can be expressed as

\begin{equation}
\label{eq:car1sol}
\epsilon (t)=e^{-\alpha_0(t-s)}\epsilon(s)+\frac{\beta}{\alpha_0}(1-e^{-\alpha_0(t-s)})+e^{-\alpha_0 t}(I(t)-I(s))
\end{equation}

\noindent
or equivalently

\begin{equation}
\epsilon (t)-\frac{\beta}{\alpha_0}=e^{-\alpha_0(t-s)}(\epsilon(s)-\frac{\beta}{\alpha_0})+e^{-\alpha_0 t}(I(t)-I(s))
\end{equation}

\section{Proof of Theorem 1}
\label{sec:teo1}
For a given positive integer $n$ we can write
\begin{eqnarray*}
y_{t_j}=\phi^{t_j-t_{j-n}} \, y_{t_{j-n}} +  \sigma  \sum_{k=0}^{n-1} \phi^{t_j-t_{j-k}} \, \sqrt{1-\phi^{2(t_{j-k}-t_{j-k-1})}}  \, \varepsilon_{t_{j-k}},
\end{eqnarray*}
Notice that under the assumptions of the theorem the first term converges to zero in probability. On the other hand, we have that
\begin{eqnarray*}
\phi^{2(t_j-t_{j-k})} \leq k^{\alpha}, 
\end{eqnarray*}
where 
\begin{eqnarray*}
\alpha = C \log \phi^2
\end{eqnarray*}
Consequently,
\begin{eqnarray*}
\sum_{k=0}^{\infty} \phi^{2(t_j-t_{j-k})} \leq \sum_{k=0}^{\infty}  k^{\alpha} <\infty,
\end{eqnarray*}
since $\alpha<-1$ by assumption. Thus, the expression
\begin{eqnarray}
\label{eq:sol}
y_{t_j}=  \sigma  \sum_{k=0}^{\infty} \phi^{t_j-t_{j-k}} \, \sqrt{1-\phi^{2(t_{j-k}-t_{j-k-1})}}  \, \varepsilon_{t_{j-k}}
\end{eqnarray}
corresponds to a measurable transformation of the independent and identically distributed (i.i.d.) sequence $\{\varepsilon_{t_j}\}$. Therefore, due to
Theorem~1.7 of  \cite{Palma_07}, the sequence $\{y_{t_j}\}$ is stationary and ergodic.\\

Further, it is straightforward to see that the equation \eqref{eq:sol} is a solution to the process defined by  (\ref{Model}). This can be shown by plugging-in $y_{t_{j-1}}$, as defined in (\ref{eq:sol}), into the right-side of equation   (\ref{Model}). After some arithmetic one gets to $y_{t_{j}}$, showing that  (\ref{eq:sol}) is indeed a solution to the process defined by (\ref{Model}). $\Box$
\vspace{0.2in}

\section{Proof of Lemma 1}
\label{sec:lem1}
It follows from Section 8.8 of \cite{Brockwell_1991}. Observe that  $t_j-t_{j-n}$ satisfies the condition of  Theorem 1, consequently, given that $0< \phi<1$, the process $\{y_{t_j}\}$ is stationary and ergodic. Furthermore, the process satisfies  the equation:
$y_{t_j}=\phi^h \, y_{t_{j-1}} +  \eta_{t_j}$,
where $ \eta_{t_j}$ is a white noise sequence with variance
$\sigma_{\eta}^2 = \sigma^2 (1-\phi^{2h})$. 
Consider the transformation $\theta=\phi^h$. Thus, an application of  Brockwell and Davis (1991, p259) yields $\sqrt{n} \, (\widehat{\theta}_n-\theta) \to \rm{N}(0,\sigma_{\theta}^2)$,
as $n \to \infty$,  where $\sigma_{\theta}^2 =1-\theta^2$. Therefore, by defining $g(\theta)=\theta^{1/h}$ we have that $\phi=\theta^{1/h}$
and then by an application of the continuous mapping theorem we conclude that
$\sqrt{n} \, ( g(\widehat{\theta}_n)-g(\theta)) \to \rm{N}(0,\sigma_{\theta}^2 \, [g'(\theta)]^2)$,
as $n \to \infty$. But, 
\begin{equation*}
\sigma_{\theta}^2 [g'(\theta)]^2 = \frac{1-\phi^{2h}}{h^2 \,\phi^{2h-2}},
\end{equation*}
which completes the proof. $\Box$

\bsp	
\label{lastpage}
\end{document}